\documentclass[aps,prb,twocolumn,groupedaddress,floatfix]{revtex4-1}
\usepackage[]{graphicx}
\usepackage{amssymb,amsfonts,amsmath}
\usepackage[version=3]{mhchem}
\newcommand{\un}[1]{\,\textup{#1}}
\begin{document}
\title{Water confined in self-assembled ionic surfactants nano-structures}
\author{Samuel Hanot}
\email[]{hanot@ill.fr}
\affiliation{Institut Laue-Langevin - 6 rue Jules Horowitz, BP 156, 38042 Grenoble, France}
\author{Sandrine Lyonnard}
\email[]{sandrine.lyonnard@cea.fr}
\author{Stefano Mossa}
\email[]{stefano.mossa@cea.fr}
\affiliation{Univ. Grenoble Alpes, INAC-SPRAM, F-38000 Grenoble, France}
\affiliation{CNRS, INAC-SPRAM, F-38000 Grenoble, France}
\affiliation{CEA, INAC-SPRAM, F-38000 Grenoble, France}
\begin{abstract}
We present a coarse-grained model for ionic surfactants in explicit aqueous solutions, and study by computer simulation both the impact of water content on the morphology of the system, and the consequent effect of the formed interfaces on the structural features of the adsorbed fluid. On increasing the hydration level at ambient conditions, the model exhibits a series of three distinct phases: lamellar, cylindrical and micellar. We characterize the different structures in terms of diffraction patterns and neutron scattering static structure factors. We demonstrate that the rate of variation of the nano-metric sizes of the self-assembled water domains shows peculiar changes in the different phases. We also analyse in depth the structure of the water/confining matrix interfaces, the implications of their tunable degree of curvature, and the properties of water molecules in the different restricted environments. Finally, we discuss our results compared to experimental data and their impact on a wide range of important scientific and technological domains, where the behavior of water at the interface with soft materials is crucial.
\end{abstract}
\date{\today}
\maketitle
\section{Introduction}
\label{sect:intro}
Surfactants are molecules that contain hydrophilic and hydrophobic groups~\cite{jones2002soft}. This peculiar amphiphilic character is the origin of their outstanding properties and widespread employment for many practical applications, ranging from biological systems to microelectronics industry~\cite{schramm20032}. When mixed with water, surfactants self-organize in order to minimize interfacial energies. This self-assembly process is driven by a subtle interplay of attractive and repulsive interactions, leading to the formation of a rich morphology of hydration-dependent structures~\cite{jones2002soft}. At high water content surfactant micelles are formed, with hydrophilic segments facing the aqueous medium. Upon decreasing hydration, various phases can be observed, with a typical organization of elongated aggregates into cubic and hexagonal packings. Eventually, lamellar phases are stabilized at very high surfactant concentrations. The variable complexity of the observed phase diagrams has been shown to mainly depend on the nature of the hydrophobic tails and the hydrophilic heads~\cite{boek2002molecular}. In {\em ionic} surfactants a negative (or positive) charge is located on the hydrophilic head, the total charge of the system being neutralized by the presence of counter-ions. The features of these materials are even more complex than those of non-ionic surfactants, due to the additional long-range coulombic interactions~\cite{faul2003ionic}. If the solvent used is polar (as water), it acts as a dielectric medium effectively screening the coulombic interactions. The presence of charges obviously influences the molecular interactions and, therefore, surfactant aggregation, generating a range of novel macroscopic properties~\cite{schramm20032}. 

Design of new architectures and consequent functionalities for advanced applications motivates in-depth studies of the structure/activity interplay in both non-ionic and ionic surfactants. This implies the use of molecular-level investigations to unveil the morphologies of self-assembled objects, provide insight into the static and dynamical properties of the solvent, and clarify their relation with the structure of the surfactant matrix. Very advanced tools, including spectroscopy and scattering techniques~\cite{zemb2002neutrons}, and computer simulation~\cite{shelley2000computer} are therefore increasingly needed. 

Also, surfactants provide well suited model systems for fundamental investigation of the effect of confinement in {\em soft} hydrophobic environments, which is a crucial issue in materials and environmental science and in biology. In general, transport properties of solvent molecules or solvated ions are highly influenced by confinement at the nano-scale. Size, shape and connectivity of the confining matrix, together with the nature of the charged interfaces and the interplay among hydrophobic, electrostatic and Van der Waals forces, have been shown to significantly alter the fluid properties compared to those of the bulk~\cite{alcoutlabi2005effects,frick2005inelastic,alba2006effects,rasaiah2008water,zanotti2012nanometric,perkin2013soft}. Most part of these studies has focused on confinement in hard matrices but, more recently, the interest in soft confinement, where boundaries mobility cannot be neglected, has substantially grown~\cite{wang2004intramicellar,hunter2014boundary}. Much more complex ionomer membranes, like Nafion, have also been extensively scrutinized~\cite{mauritz2004state}. Yet, relatively scarce studies have been devoted to the understanding of water structure and dynamics in ionic surfactants, in different hydration conditions~\cite{lyonnard2010perfluorinated,berrod2014}.  

Here we introduce and characterize a numerical model for ionic surfactants in solution, considering a molecular model for water (explicit solvent). Our goal is to fill the gap between coarse-grained approaches on one hand and fully atomistic resolution on the other. We therefore trade an acceptable degradation of spatial resolution for the surfactants molecules for a complete molecular description of the fluid, keeping the possibility of following the evolution of important degrees of freedom
and grasping both the details of the nano-structure of water domains (1-15\un{\AA}) and the macroscopic ordering of surfactant aggregates (20-100\un{\AA}).
\begin{figure*}[t]
\centering
\includegraphics[width=0.99\textwidth]{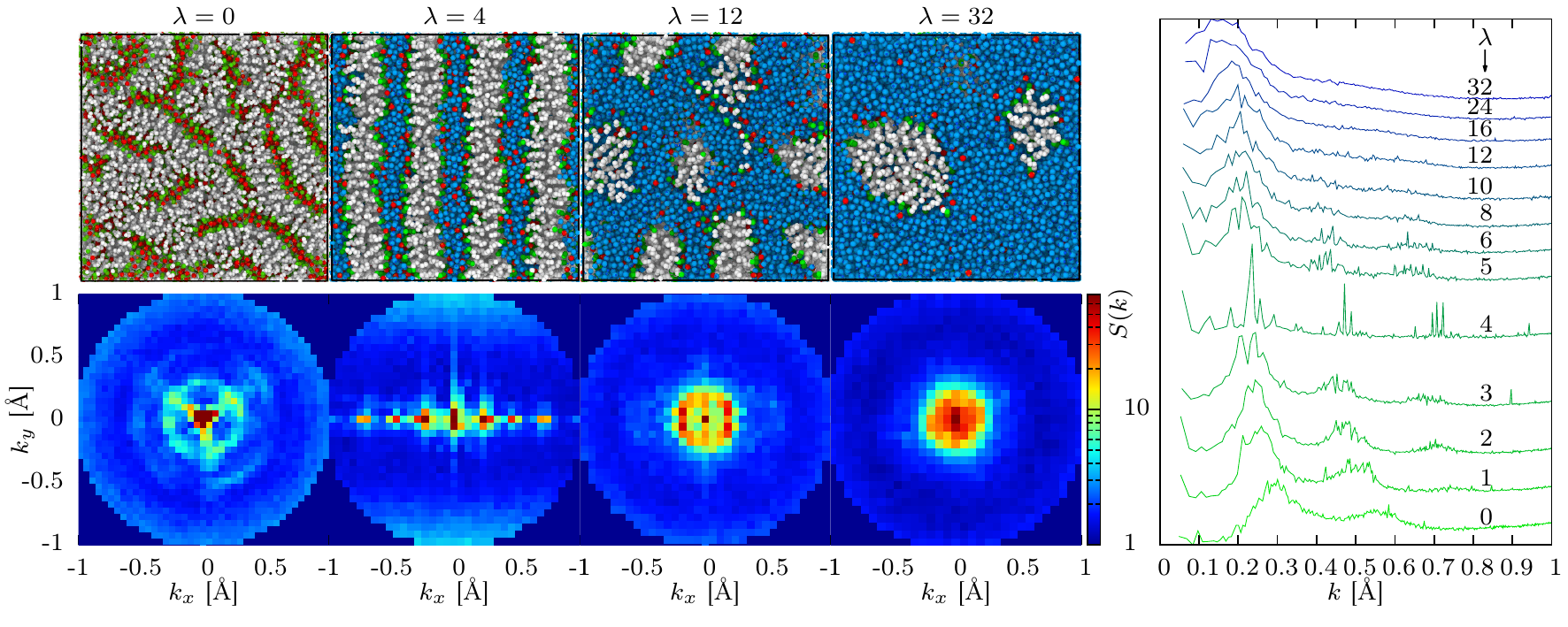}
\caption{
{\em Left, Top:} Snapshots of the self-assembled surfactant phases at the indicated values of the hydration level, $\lambda$, increasing from top to bottom. The typical simulation box contains a number of interacting units, $N$, in the range $80\div 90\times 10^3$ and has an edge length of $\simeq 100\un{\AA}$. Periodic boundary conditions are used in all directions. Hydrophobic segments of the surfactants are in white, sulfonated heads in green, water molecules in blue, hydronium complexes in red. {\em Left, Bottom:} Diffraction patterns for the corresponding phases on the top, calculated as discussed in the SI. The intensity is visualized as a color-map in the $(k_x,k_y)-$plane. {\em Right:} Neutron scattering static structure factors, $S(k)$, at all investigated values of $\lambda$. Curves have been shifted arbitrarily for clarity.\label{fig:snapshots}}
\end{figure*}
We have selected a macro-molecular structure, similar to the perfluorooctanesulfonic acid (PFOSA)~\cite{kissa2001fluorinated} which features a highly hydrophobic perfluorinated chain, while polarity is provided by the sulfonic acid group. The combination of a perfluorinated backbone and superacid terminal functions favours the formation of a sharp interface between hydrophobic aggregates and the aqueous phase, thus making PFOSA an ideal model system for investigating both surfactant self-assembly and solvent properties. This also enhances the possibility of studying in unprecedented details, to the best of our knowledge, the self-organization of hydrophobic/hydrophilic interfaces, and their impact on the adsorbed fluid features.
\section{Results}
\label{sect: results}
Without going into the details of the chemical structure, our ionic surfactants solution is a three-components mixture: the surfactant macro-molecules (Su), the polar solvent (water, W), and the solvated (counter-)ions (hydronium, H). The physics of the material is controlled by the interplay of: {\em i)} the energy of the interfaces formed between the amphiphilic aggregates and the fluid phase, and {\em ii)} the electrostatic interactions among the charged hydrophilic heads and the solvated cations. Developing a good model is tantamount with tuning the relative weights of these two terms, stabilizing the attended phases at different water contents. 

We have chosen a united-atoms representation for the surfactant molecule (Fig.~\ref{fig:C8} in the Methods section). This is inspired by the model of Allahyarov and Taylor~\cite{allahyarov2008simulation} for the side-chain of Nafion. We represent the hydrophobic section of the molecule with a series of $7$ Lennard-Jones (LJ) neutral beads, each representing an entire \ce{CF2} group. Similarly, the head group is schematized by two charged LJ beads, one for the sulfur atom $S$ and one for the \ce{O3} group, with a total charge $q=-e$. This also allows to associate to the charged head a realistic value for the dipole~\cite{allahyarov2008simulation} and imposes that all acid groups are dissociated. The mass of each bead corresponds to the sum of the atomic masses pertaining to the bead. Details about the beads are given in Table~\ref{table:species} in the Methods section, while functional forms and parameters for the bonded interactions are given in Table~\ref{table:bonded} (Methods). In order to enhance the formation of well ordered lamellar phases at low hydration, we have re-optimized the original values~\cite{allahyarov2008simulation} of the parameters controlling the non-bonded interactions. Our new parametrization of the non-bonded interaction terms is given in Table~\ref{table:non-bonded} in the Methods section. The simulation box is initialized by considering a variable number of surfactants, together with the amount of hydronium complexes needed for neutralizing the total charge. The hydration level of the investigated systems is encoded in the parameter $\lambda$, i.e., the number of water molecules per surfactant. Additional details of the simulations are given in Methods section. We stress that the structures produced are the result of a very efficient {\em unbiased} self-assembly, starting from completely disordered system configurations. The initial seed of the final phase formed quite quickly, on time-scales of the order of a few nanoseconds, followed by slow relaxation, to locally optimize mechanical stress. 

The discussion of the phase diagram of our model generally conforms to that of concentrated amphiphile solutions discussed in many textbooks~\cite{jones2002soft} for increasing surfactant concentration (here for decreasing $\lambda)$. In Fig.~\ref{fig:snapshots} (left top) we present typical snapshots of the system at the indicated hydration values. Starting from the highly hydrated systems ($\lambda\geq 16$), at the lowest surfactant concentration, we find a micellar solution, with aggregates of different sizes and elongations, concentrating the hydrophobic beads at the center and exposing the sulfonate groups to water. By increasing surfactants concentration ($12 < \lambda < 16$) repulsion between aggregates becomes significant, and micelles start to show an increasingly elongated character. This can be understood by the fact that, in the present high density condition, cylinders pack in space more efficiently than spheres. This process controls the transition for $8 \le \lambda \le 12$ to a cylindrical structure with some degree of closed-packed order (see $\lambda=12$ in Fig.~\ref{fig:snapshots}), although it does not express as a fully developed hexagonal phase. In the intermediate range $6 \le \lambda < 8$ a new change occurs, with cylindrical aggregates merging and starting to transform in flat bilayers. Eventually, for $\lambda \le 4$, surfactants arrange in extremely well-ordered lamellar phases, spanning the entire simulation box, with increasingly thin ionic channels intercalating the lamellae (see discussion below). In the case $\lambda=0$, notwithstanding the absence of water, the presence of the hydronium complexes is sufficient to drive phase separation with a local lamellar structure. The structure is however not completely ordered, with hydrophobic domains organized with different orientations and the presence of a few disconnected ionic domains. Altogether, these data demonstrate that our model is able to grasp the overall phase behavior of sulfonated ionic surfactants~\cite{lyonnard2010perfluorinated}, in a wide range of hydration conditions. In all cases a sharp phase separation is evident, with the charged sulfonic heads decorating the interface between the adsorbed fluid and the hydrophobic sections of the surfactants. Hydroniums condensate at the charged interface at the lowest values of $\lambda$, while a significant fraction is solvated by water, far from the interfaces at high hydration.  

The above qualitative description of the phase diagram can be completed by characterizing the system organization via the neutron scattering static structure factors. We have calculated both the diffraction diagrams measured in experiments (color maps in Fig.~\ref{fig:snapshots}, left bottom) and the angular average, $S(k)$ (Fig.~\ref{fig:snapshots}, right), where $k$ in the wave-vector. Details about our calculations are given in the Methods section. For the almost dried system, $\lambda=0$ (Fig.~\ref{fig:snapshots}, left bottom), we observe strong Bragg peaks on three preferred directions. This is consistent with the picture of local lamellar arrangements which bend and branch, as can be seen on the corresponding simulation snapshot. At $\lambda=4$, high intensity spots, corresponding to genuine first, second and third-order Bragg peaks, appear at three non-zero positions. These are aligned along a preferential direction, which is characteristic of the lamellar arrangement with long-range order. By increasing hydration, the second and third order peaks fade, indicating a lower degree of order. At the highest hydration, the high-intensity region eventually arranges in a ring-like shape, indicating that no preferred orientational order is present in the system, as expected in the fully developed micellar phase. 

The total static structure factors, at all hydrations, are shown in Fig.~\ref{fig:snapshots} (right panel). The main features of these curves are the first peak at low-$k$, whose position corresponds to the average distance between surfactant clusters, and the higher order peaks, at larger $k$, which carry information about the arrangement of surfactants and, therefore, of the phase symmetry. We observe that, as we increase the hydration, the position of the peaks shift toward smaller wave-vectors, indicating that the distance between the surfactant aggregates increases. This is the expected effect of swelling of the system when increasing water content. Also, the highest-symmetry phase is found for $\lambda=4$, where the smectic order associated to the lamellar phase is maximum and three well-defined Bragg peaks appear. Additional information, coming from an analysis of the partial contributions $S_{\alpha\beta}(k)$ of chemical species $\alpha, \beta$, is given in the Methods section.
\section{Discussion}
\label{sect:discussion}
\begin{figure}[t]
\centering
\includegraphics[width=0.49\textwidth]{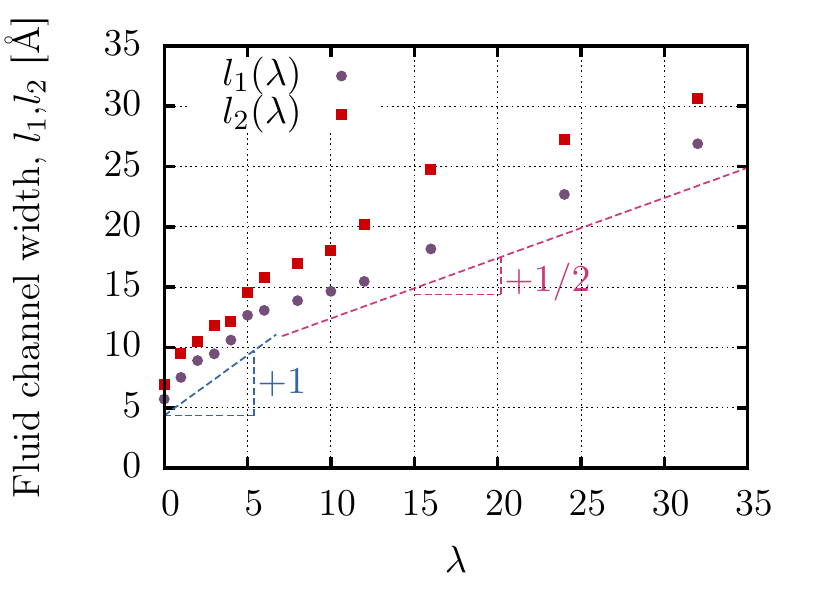}
\vspace{-0.25cm}

\caption{ $\lambda$-dependence of the size of ionic channels as calculated from the position of the first Bragg peak, $l_1$, in the static structure factors and from the direct determination of the aqueous domains extent, $l_2$, as discussed in the text. The dashed lines are guides for the eyes, of slope 1 and 0.5 at low and high hydration, respectively.\label{fig:channel_size}}
\end{figure}

Interestingly, the above {\em global} spatially-averaged static structure factors provide precise information about {\em local} geometrical properties of the confining matrices. More precisely,  we are now in the position to discuss the variation of the typical size of the ionic domains available for transport of the absorbed fluid, as a function of the hydration level. From the position $k_0(\lambda)$ of the first-order Bragg peak, we can extract $l(\lambda)=2\pi/k_0(\lambda)$, which is commonly associated to the average distance between the centers of mass of the hydrophobic aggregates. From these data we can build $l_1(\lambda)=l(\lambda)-2l_{s}(\lambda)$, where $l_s$ is the average end-to-end distance of the surfactants. Helped by inspection of the snapshot at $\lambda=4$ of Fig.~\ref{fig:snapshots}, it is easy to convince oneself that  $l_1(\lambda)$ is an {\em indirect} measure of the size of the ionic channels. Our data are shown in Fig.~\ref{fig:channel_size} and quantify the swelling behavior of the system. For $\lambda\le 6$, $l_1$ increases linearly with $\lambda$, with a slope $\simeq 1$. This corresponds to the expected swelling in the lamellar phase (see, for instance, Ref.~\cite{berrod2014}), where the distance between surfactant domains increases affinely, following the volume increase of the intercalated fluid. At higher values of $\lambda$, a clear cross-over is visible, with $l_1$ that can be interpolated by a straight line of slope $\simeq 1/2$. Rationalizing this behavior is however difficult, due to the presence of interfaces of increasing curvature and sensible disorder in the distribution of the hydrophobic aggregates.  

These data can be further strengthened by a {\em direct} independent measure of the width of the ionic channels, by computing the distribution of the distances of the fluid molecules (both water and hydronium) to the closest surfactant sulfonated head. From these distributions we have estimated the channels sizes, $l_2$, as the distance encompassing $99\%$ of the fluid molecules. Our results are shown in Fig.~\ref{fig:channel_size}. For $\lambda\le 6$ we have $l_1\simeq l_2$, as expected for highly symmetric phases. At higher values of $\lambda$, $l_2$ is always higher than, although of the same order of, $l_1$, providing an upper bound to the size of the ionic domains at each $\lambda$ in the presence of disordered curved interfaces.

\begin{figure}[b]
\centering
\includegraphics[width=0.49\textwidth]{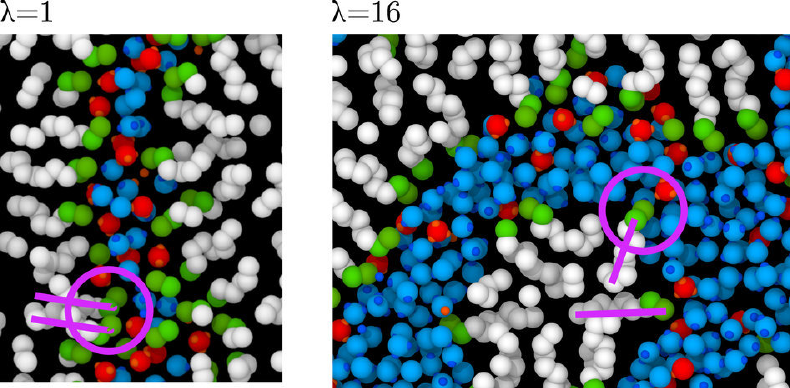}
\caption{Details of typical generated interfaces with different curvature radius, together with a schematic representation of the hydration sphere (circles) used for our calculations and the mutual orientations of adjacent surfactants (lines). The snapshot on the left corresponds to $\lambda=1$ (flat interface), to $\lambda=16$ that on the right (curved interface).\label{fig:interface}}
\end{figure}

{\em Interfaces} play a crucial role in the physics of our surfactants systems. In Fig.~\ref{fig:interface} we show local configurations of the interfaces generated at $\lambda=1$ and $16$, in the (pseudo) lamellar and micellar phases, respectively. The sketched segments indicate the relative orientations of selected adjacent surfactants, parallel (left) and at a non-zero angle (right) in the two cases. Obviously these conform to the charge distribution geometry at the interfaces, planar in one case and with a finite curvature radius in the other. We can expect the coordination features of both surfactants and fluid molecules, and therefore wetting properties at the interface, to be sensibly different in the two cases. We therefore first focus on the coordination properties of surfactants heads, in the coordination sphere indicated by the circles in Fig.~\ref{fig:interface}.

\begin{figure}[t]
\centering
\includegraphics[width=0.48\textwidth]{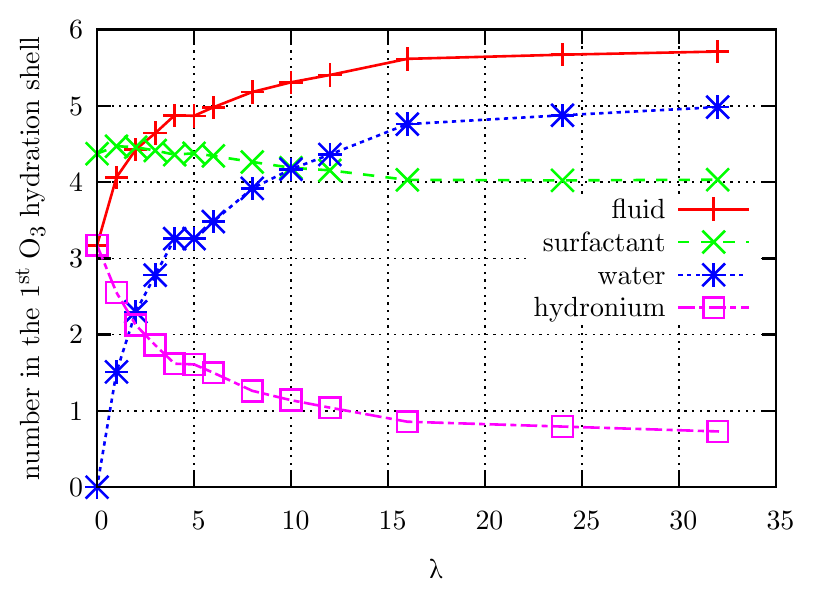}
\vspace{-0.25cm}

\includegraphics[width=0.48\textwidth]{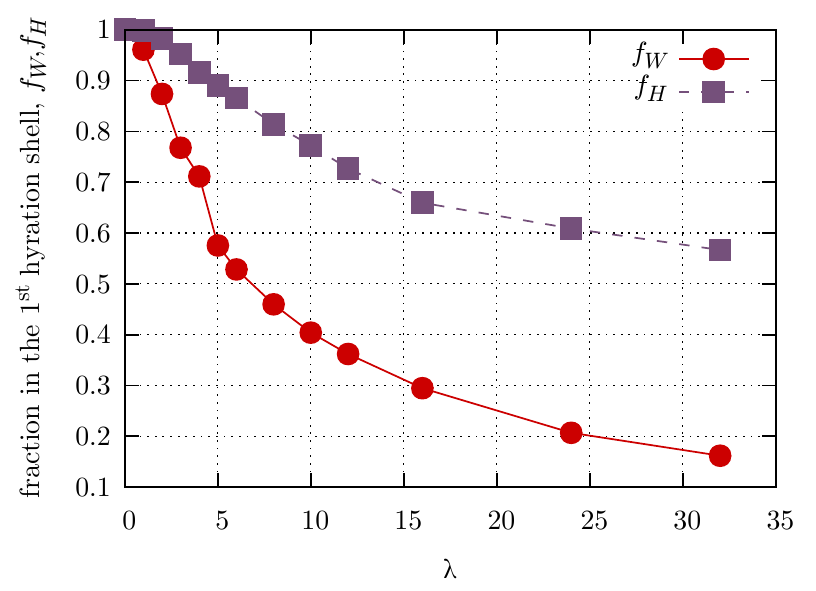}
\caption{
{\em Top:} $\lambda-$dependence of the average number of molecules of the indicated species localized into the first coordination shell of the \ce{O3} bead of the surfactant heads. {\em Bottom:} $\lambda-$dependence of the average number of water molecules and hydronium complexes localized into the first coordination shell of the \ce{O3} bead of the {\em closest} surfactant head. A detailed discussion of these data is included in the main text.\label{fig:coordnum_all}
}
\end{figure}

In Fig.~\ref{fig:coordnum_all} (top), we plot the average number of molecules of the different species comprised in the first coordination shell of the surfactant heads. This is defined as the sphere with radius $R_c=4.3\un{\AA}$, corresponding to the first minimum in the \ce{O3}-\ce{O_W} pair radial distribution functions (not shown). These data show that, for $\lambda \leq 5$, the first coordination shell of the surfactant heads contains about 4.3 atoms pertaining to surfactant molecules. In the same  hydration range, the number of fluid molecules in the first coordination shell increases from 3.1 to slightly less than 5. It is interesting to note that there is no variation of this number between $\lambda=4$ and $\lambda=5$. More precisely, at $\lambda=4$ the first coordination shell is saturated and added water molecules organize far from the interfaces, as we will see more in the details below. For $\lambda \geq 5$, we observe that the number of surfactant molecules inside the first coordination shell decreases, while it increases for fluid molecules. This is the manifestation of the lamellar-to-cylindrical phase transition. By increasing $\lambda$, the system efficiently packs more fluid molecules at the hydrophilic interface, by increasing the tilt angle between adjacent surfactant and, as a consequence, the mutual distance of the sulfonic heads. This curves the exposed interface, giving rise to increasingly well defined cylindrical structures. At $\lambda=16$, the number of both fluid and surfactant particles in the first coordination shell reach limiting values, which show very mild variation at higher degrees of hydration. This is a direct signature of the transition to the micellar phase, where spherical aggregates are dissolved in (bulk) water, with a total surface area of the interfaces almost independent of $\lambda$.

We can push further the analysis of the interfaces by counting the per/surfactant average number of water molecules (or hydronium ions) that are in the first coordination shell of their {\em nearest} hydrophilic head, ${\cal N}_{W(H)}=1/{N_s}\sum_{m \in W(H)} \left[1-\theta(z_m-R_c)\right]$, where $z_m$ is the distance of molecule $m$ to the nearest hydrophilic head. We note that this procedure eliminates the over-counting intrinsic to the above discussion, where the same fluid molecule could be considered as belonging to the first coordination shell of multiple surfactant heads. Also, by construction $0 \leq {\cal N}_H \leq 1$ and $0\leq {\cal N}_W \leq \lambda$. The quantities $f_H={\cal N}_H$ and $f_W={\cal N}_W/\lambda$ can therefore be interpreted as the fraction (over the total number) of hydronium ions and water molecules in direct contact with the interface. These data are included in Fig.~\ref{fig:coordnum_all} (bottom) and show a few interesting features. 

At very low hydration ($\lambda=1$), almost all water molecules and hydronium ions are condensed in contact with the nearest hydrophilic head, resulting in ${f}_{W(H)} \approx 1$. This is evidence of a phase where all fluid molecules are localized within the first coordination shell of their nearest surfactant heads, in very thin ionic layers intercalated between the surfactant planes. For $1\leq \lambda \leq 4$ we observe an almost linear decrease of $f_W$, which however still keeps a quite high value ($\simeq 70\%$) at $\lambda=4$. Indeed, in that hydration range half of the width of the fluid channel ($l_1(\lambda)$ in Fig.~\ref{fig:channel_size}) is smaller than or comparable to the radius of the coordination shell, meaning that most part of the added water molecules will be placed at the interface. In the same range of $\lambda$, $f_H$ only decreases by a mere $10\%$, indicating however that, notwithstanding the strong Colombic interactions which keep hydronium ions strongly bound to the sulfonate groups, there is still the possibility for a water molecule to substitute a cation at the interface. 

At $\lambda=5$, where the lamellar order is still present, the fluid channels are no longer spanned by the coordination regions of facing surfactant heads, which are saturated, and additional fluid molecules are therefore placed outside the interfaces. For $\lambda>5$, the sequence of phase transitions described before occurs, implying the progressive curvature of the interfaces. This effect increases the interfaces area available for fluid adsorption compared to that present in the lamellar phase and controls the reduction of $f_W$ upon hydration, explaining the clear cross-over to a less-than-linear behaviour at high $\lambda$. Eventually, at very high hydration, only about $10\%$ of the total amount of water molecules is still in contact with the interfaces, the vast majority having a prominent bulk-like character. This is at variance with the case of the hydronium ions, which even in very high hydration conditions substantially keep their interfacial character, with $\simeq 50\%$ of the entire population still bounded at the charged interfaces.   

\begin{figure}[t]
\centering
\includegraphics[width=0.48\textwidth]{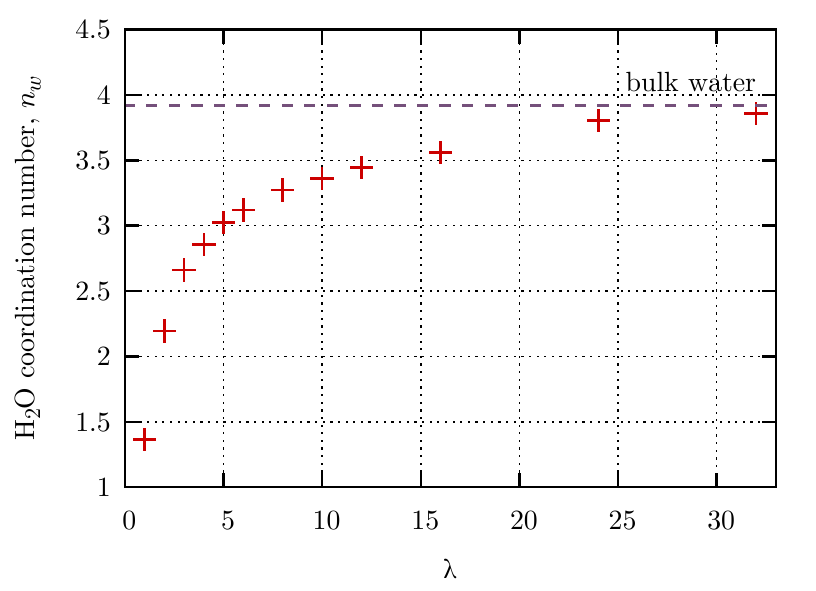}
\vspace{-0.25cm}

\includegraphics[width=0.48\textwidth]{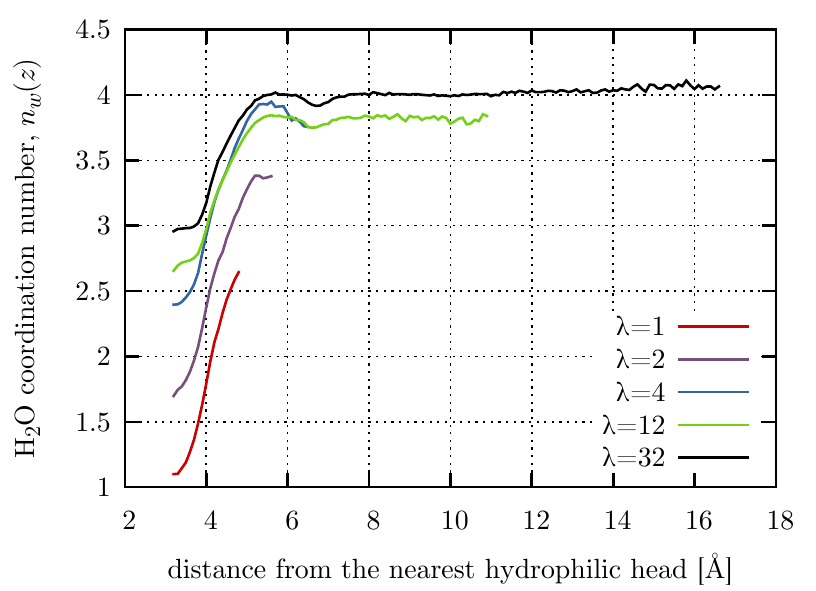}
\caption{
{\em Top:} $\lambda-$dependence of the average water-water coordination number, $n_W(\lambda)$. The dashed line indicates the values $\simeq 4$ for pure bulk water. {\em Bottom:} Space-dependent water-water coordination number, $n_W(z,\lambda)$, as a function of the distance, $z$, of the molecule from the closest hydrophilic head. We only show our results for the indicated values of $\lambda$, for clarity.\label{fig:water-coord}
}
\end{figure}

We now clarify the effect of the above subtle interface/bulk character of water molecules in different regions of the ionic domains on the coordination properties of the absorbed fluid itself. We focus, in particular, on the water-water average coordination number and its variation at different distances from the interfaces. In Fig.~\ref{fig:water-coord} (top) we plot the average coordination number of water molecules, $n_W(\lambda)$ at all considered hydration levels. The value for bulk water $\simeq 4$ is shown for reference. These data have been calculated with the usual technique of integrating the pair radial distribution function, $g_{\ce{O_W}-\ce{O_W}}(r)$, in the range $[0:R_C]$, with $R_c=3.16 \un{\AA}$ the position of its first minimum. At high hydration, we recover the value of bulk water $n_W\simeq 4$ (dashed line). By decreasing $\lambda$, $n_W$ steadily decreases to $n_W\simeq 3$ at $\lambda=6$, at the boundary with the lamellar phases. Next, due to the increasing extent of affine confinement, the observed decrease is much steeper, and $n_W$ reaches a minimum value $\simeq 1.2$ at $\lambda=1$. This is consistent with the general picture discussed above, where water molecules, hydronium ions and sulfonate groups are intercalated in the ionic channels, organizing in almost mono-planar configurations.

Our simulations allow us to access even more detailed information, e.~g., the space-dependent coordination number of water molecules, in different regions of the ionic channels. In Fig.~\ref{fig:water-coord} (bottom), we show the average coordination number of molecules at distance $z$ from the closest sulfonate group, $n_W(z,\lambda)$. Obviously, $\int_{0}^{z_{max}} dz \; n_W(z,\lambda)\; \rho(z,\lambda)=n_W(\lambda)$, where $\rho(z,\lambda)=\sum_{i=1,N_W}\delta(z_i-z)/N_W$ is the number density of water molecules and $z_i$ the distance of water molecule $i$ from the closest surfactant head. These data show that, at all $\lambda$'s, $n_w(z,\lambda)$ steadily increases from a highly suppressed value at $z_{min}\simeq 3\un{\AA}$ in contact with the interface ($z_{min}$ is mainly controlled by the value of $\sigma_{\ce{O_3} \ce{O_W}}$, see Table~3 in the SI) to a maximum at the $\lambda$-dependent $z_{max}$, at the center of the channels ($z_{max}\simeq\l_2/2$, at all hydrations). The overall variation of $n_W(z,\lambda)$ with the distance depends on $\lambda$, decreasing from a factor of $\simeq 2.5$ at $\lambda=1$ to $\simeq 1.3$ for $\lambda=32$. We conclude our discussion on this point by observing that the details of the curve corresponding to the highest hydration level allow us to highlight the presence of three different regions inside the ionic domains, determining the most prominent character of the confined water molecules: {\em i)} {\em bulk-like} at large distances, $z\ge 8\un{\AA}$, in the center of the ionic domains and far from any boundaries; {\em ii)} {\em intermediate}, corresponding to distances $z\simeq 6\un{\AA}$, where the shallow spatial modulation of $n_W$ is due to the well-known layering of water molecules; and {\em iii)} {\em interface-limited}, where the extent of nano-confinement is so high to completely destroy any resemblance of the hydrated environment with the one in the bulk phase. 
\section{Conclusions}
\label{sect:conclusions}
We have developed a coarse-grained model of ionic surfactants in solution, with explicit water molecules and hydronium complexes as counterions. Extensive Molecular Dynamics simulations have been performed, over a wide range of water contents ranging from the almost dried condition, where counter-ions only are present, to highly hydrated states. We have observed very efficient system self-assembly at all hydration levels, solely driven by the forces generated by the chosen interaction potential, starting from disordered initial configurations with no initial bias toward any phase symmetry. The generated phase behaviour has been demonstrated to nicely mimic that of sulfonated ionic surfactants, at ambient conditions, encompassing lamellar, cylindrical and micellar structural organization. In all cases we have observed sharp charged interfaces separating the aqueous domains from the confining soft matrix, formed by the hydrophobic segments of the surfactants. Based on a very detailed analysis, the confinement sizes associated to the ionic domains have been found to increase with hydration, in the range $5\div 25\un{\AA}$. A transition between affine swelling and a more complex behaviour has been found at length scales of the order of $14\un{\AA}$. 

Neutron scattering static structure factors were also calculated, similar to those determined in experiments. Neutron and X-rays scattering measurements on PFOSA molecules of size comparable to that of our model were reported recently~\cite{lyonnard2010perfluorinated,berrod2014}. A nice agreement is found between our data and the experimental results, as the boundaries of the different phases, typical average inter-aggregate correlation distances and overall swelling behaviour are similar. Our model is therefore capable of realistically reproducing the PFSOA surfactant solution behaviour. This suggests that it can be used as an {\em in-silico} partner to the experimental study of a variety of surfactants. For instance, we could easily tune the surfactant length and generate different commercially available materials, as PFHSA (6 carbons) or PFBuSA (4 carbons)~\cite{kissa2001fluorinated}. The phase behaviour of these systems being quite universal, we expect limited variations compared to the present case, with the modified length of the surfactants probably controlling the hydration-level location of the different phase boundaries. 

In the same context, our model could also be valuably applied to generate and investigate {\em polymeric} systems, in particular perfluorosulfonic acid polymer electrolyte membranes. More specifically, the quite ordered PFOSA surfactant phases were recently considered as a facilitated playground for clarifying microstructure and transport properties of nanoscopic phase-separated ionomers~\cite{lyonnard2010perfluorinated,berrod2014}. In particular, the PFOSA molecule is very similar to the pendant side-chain of Nafion, the most investigated of the PFSA polymer electrolyte membranes used in fuel cells and electrolysers (Ref.~\cite{kreuer2013ion}, and references therein). Unfortunately, even the most modern scattering techniques provide averaged spatial information only~\cite{gebel2005neutron}, which does not allow to form a comprehensive consistent picture of both the organization of these extremely complex materials at the nanoscale, and the consequent impact on transport properties. Hydrated sulfonated ionic surfactants have therefore been proposed as model systems for the {\em local} organization of ionomers, to quantify, among other issues, the effects of well-controlled confining geometries on proton mobility~\cite{berrod2014}.

Based on these facts, we can conclude that with the present model we are in the position to unify in a single bottom/up computational framework the entire vast class of perfluorinated materials. Indeed, we have demonstrated here that our re-optimized force-field correctly encodes the phase behaviour and the most important structural features of a PFOSA system. Due to the similarities of the latter with the side chain of Nafion, we can employ this structure as the fundamental building-block to upscale our description to the entire ionomer, by grafting it to a strongly hydrophobic polymer backbone with tuned mechanical properties ({\em i.e.}, persistence length). This is of course in the same spirit of the original work of Ref.~\cite{allahyarov2008simulation}, but with the crucial difference that we can now count on a model with physically-sound behaviour at all length scales, ranging from local lamellar-like structures to long-range organization of ionic and hydrophobic domains.\footnote{We note that Ref.~\cite{sunda2013molecular} contains an extremely detailed all-atoms simulation study of the side chain pendants of three perfluorosulfonic acid polymer electrolyte membranes, characterized by different structures. Structural organization was investigated in details, together with some transport properties, at different degrees of hydration. Interestingly, the Authors completely overlooked the possibility to comment on similarities with ionic surfactant phases.} We also note that our approach naturally provides all ingredients for generating a virtually infinite range of composite materials, obtained by doping the ionomer with elongated charged macro-molecules, including ionic liquids~\cite{sood2012proton,lu2014chemical}.

Our results go well beyond the particular (sulfonic) moiety considered. More generally, our model can be seen as a powerful tool for generating efficient self-assembly of soft interfaces with a controlled degree of curvature. Recent fundamental work~\cite{chandler2005interfaces,willard2014molecular,chandler2011lectures} has focused on assembly driven by hydrophobic forces, even in the presence of hydrophilic units, as it is the case here. It has been shown that the properties of water molecules at the interface with extended molecular aggregates are certainly controlled by the nature of the interactions between water molecules and the aggregates, but a crucial role is played by the extent of the surface occupied by the aggregates itself. In particular, the associated curvature radius strongly affects the wetting features, or the hydrogen bonding formation. In the limiting case of purely hydrophobic aggregates of large available surface, water molecules at the interface even acquire a gas-like character, with a strongly suppressed interfacial density~\cite{willard2014molecular} and a number of formed hydrogen bonds significantly lower than that in the bulk~\cite{chandler2005interfaces}. In the present case, the situation is sensibly more complex, with the presence of hydrophilic interactions between surfactant heads and water molecules, and the possibility to form additional hydrogen bonds with both the hydronium complexes condensed at the interfaces and the fully dissociated sulfonated groups. However, our data also point to non-trivial modifications of the character of water molecules, in the presence of interfaces with different degrees of curvature. In particular, the degradation of both average and space-dependent (at different distances from the interfaces) coordination properties are expected to impact structure and life-times associated to the hydrogen bonds network and, in general, transport properties of the adsorbed fluid itself. We are convinced that our work provides a well-designed general foothold to attack all these issues, which are of paramount importance in modern science, ranging from soft synthetic materials to systems of biological interest~\cite{bizzarri2002molecular,yamamoto2014origin}.
\section{Methods}
\label{sect:methods}
{\bf State of the art.} Molecular dynamics simulations at the mesoscale are the most indicated tool to address the issues discussed in the main text. It is interesting to note that only limited surfactants phase diagram regions have been generally considered in the literature, focusing particularly on micellar phases~\cite{Jorge2008,Jusufi2009,Sayyed-Ahmad2010,Sanders2010}, in high hydration conditions. These works mostly report phase diagrams and basic structural details, in the form of pair distribution functions and typical sizes of surfactant clusters. Very limited attention is normally devoted to the features of the adsorbed fluid. Furthermore, most part of the existing numerical investigations are somehow exceedingly focused on the chemical details of the materials or, in contrast, they miss important details in the description of the solvent~\cite{Shinoda2008,Klein2008,bhattacharya2001self,Shinoda2008a}, relying on heavily coarse-grained models. Obviously, increasing spatial resolution to the atomic level for the surfactant molecules would necessitate unreasonably extended computing resources.On the other hand, neglecting some details of the water molecule has an important influence on the dynamics of the simulated systems, and hampers the possibility of following the evolution of important degrees of freedom. 
\begin{figure}[b]
\centering
\includegraphics[width=0.49\textwidth]{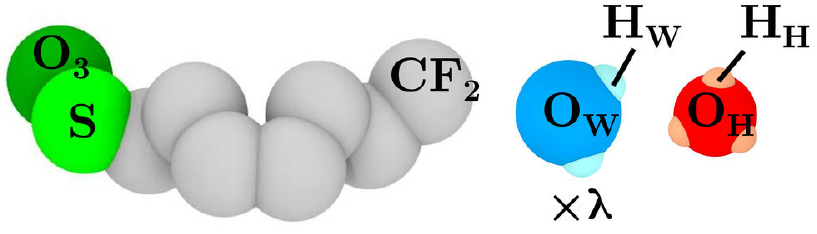}
\caption{
Sketches of the geometry used for the molecules considered in this study: ionic surfactant (left), water (center), hydronium (right). The geometry of the surfactant molecule is the one introduced in Ref.~\cite{allahyarov2008simulation}.
}
\label{fig:C8}
\end{figure}
\begin{table}[t]
\caption{
Numerical values of quantities associated to the different interaction units of species $\alpha$, including united-atoms beads and atoms: mass, $m_\alpha$, charge, $q_\alpha$, and neutron scattering length, $b_\alpha$. Masses and scattering lengths of the united-atoms beads are the sum of the values of those pertaining to the constituent atoms.
}   
\label{table:species}
\vspace{0.5cm}

\centering
\begin{tabular}{l|ccc}
\hline
$\alpha$      &	$m_\alpha$ [g mol$^{-1}$]	&	$q_\alpha$ [$e$]    &   $b_\alpha$ [fm]\\
\hline\hline
\ce{CF2}			&	50.0						&	0			    &   17.954  \\
S	 			&	32.0						&	1.1			    &   2.847   \\
\ce{O3}			&	48.0						&	-2.1		    &   17.409 \\
\ce{O_W}		&	15.99940					&	-0.8476		    &   5.803   \\
\ce{H_W}		&	1.0080						&	0.4238		    &   -3.7390 \\
\ce{O_H}	&	15.99940					&	-0.2480		    &   5.803   \\
\ce{H_H}	&	1.0080						&	0.4160		    &   -3.7390 \\
\hline    
\end{tabular}     
\end{table}
\begin{figure}[b]
\centering
\includegraphics[width=0.49\textwidth]{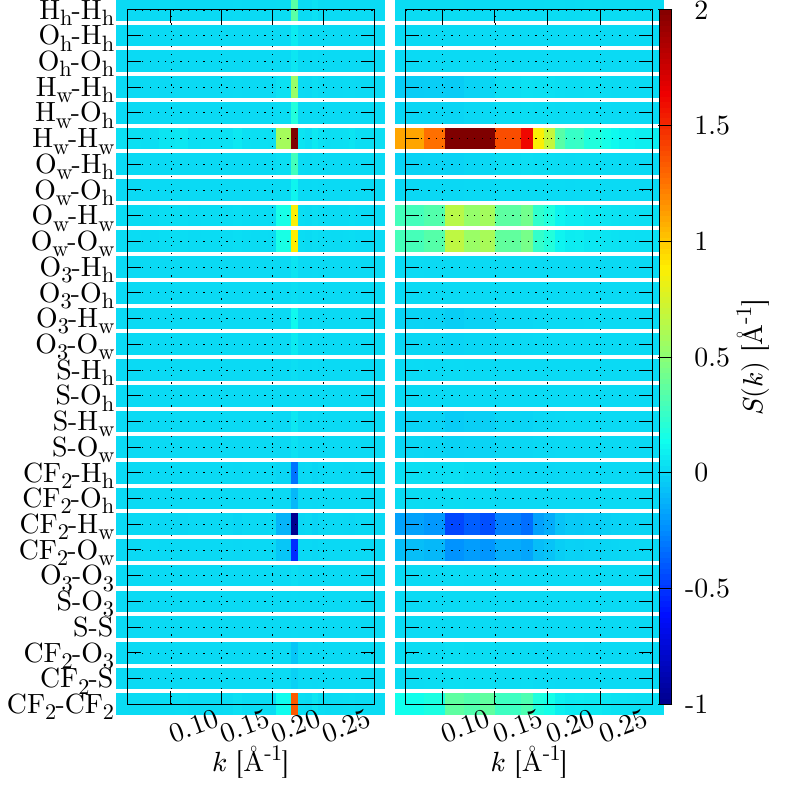}
\caption{
Partial static structure factors, $S_{\alpha\beta}(k)$. We have considered all interacting pairs $\alpha,\beta$ with $\alpha\le\beta$, for a total of $28$ different contribution. We show the results for $\lambda=4$ (left) and $\lambda=32$ (right).
}
\label{fig:skab}
\end{figure}
\begin{table*}[t]
\caption{
Parameters for the intramolecular potential~\cite{allahyarov2008simulation}. Details are given in the text.
}     
\label{table:bonded}
\vspace{0.5cm}
\centering
\begin{tabular}{l | c l|  c l | c c c }
& \multicolumn{2}{c}{Bonds: $k_B(r-r_0)^2$} &\multicolumn{2}{c}{Angles: $k_A(\theta-\theta_0)^2$}&\multicolumn{3}{c}{Dihedrals: $k_D[1+d\cos(n\phi)]$}\\
\hline
Molecule& $k_B$ [kcal mole$^{-1}$ \AA$^{-2}$]	& $r_0$ [\AA] & $k_A$ [kcal mole$^{-1}$
deg$^{-2}$]	& $\theta_0$ [deg] & $k_D$	[kcal mole$^{-1}$ deg$^{-2}$]	
&	d	&	n\\
\hline\hline
Su	&	350.0	&	1.540 & 60.0	&	110.0	&	3.1	&	-1	& 3\\
W		&	450.0	&	1.0	  & 55.0	&	109.47  & --&--\\	
H	&	450.0	&	0.973 & 55.0	&	111.6  & --& --\\
\hline
\end{tabular}
\end{table*}
\begin{table}[b]
\caption{
Parameters for the non-bonded Lennard Jones interaction potentials. Details are given in the text.
}   
\label{table:non-bonded}
\vspace{0.5cm}
\centering
\begin{tabular}{l l| l l r}
\hline
$\alpha$ & $\beta$ & $\epsilon_{\alpha\beta}$ [kcal/mole]	&	$\sigma_{\alpha\beta}$ [\AA]	& $r_{c,\alpha\beta}$ [\AA] \\
\hline\hline
\ce{CF2}		    &	\ce{CF2}          &	0.20	&		3.93		&		11.79\\
\ce{CF2}		    &	S	            &	0.05	&		3.93		&		11.79\\
\ce{CF2}		    &	\ce{O3}           &	0.05	&		3.93		&		11.79\\
\ce{CF2}		    &	\ce{O_W}        &	0.4		&		3.93		&		4.41\\
\ce{CF2}		    &	\ce{O_H}    &	0.4		&		3.93		&		4.41\\
S       	    &	S	            &	0.15	&		3.93		&		11.79\\
S       	    &	\ce{O3}		    &	0.15	&		3.93		&		11.79\\
S       	    &	\ce{O_W}     &	0.4		&		3.93		&		11.79\\
S       	    &	\ce{O_H}   &	0.4		&		3.93		&		11.79\\
\ce{O3}		    &	\ce{O3}		    &	0.15	&		3.93		&		11.79\\
\ce{O3}		    &	\ce{O_W}     &	0.4		&		3.93		&		11.79\\
\ce{O3}		    &	\ce{O_H}   &	0.4		&		3.93		&		11.79\\
\ce{O_W}    &	\ce{O_W}    &	0.1554	&		3.1655		&		9.50\\
\ce{O_H}    &	\ce{O_H}    &	0.2740	&		2.9		    &		8.70\\
\ce{H_W}    &	*      &	0.0		&		0.0		    &		0.00\\
\ce{H_H}  &	*   &	0.0		&		0.0		    &		0.00\\
\hline
\end{tabular}
\end{table}

{\bf The model.} As mentioned in the main text, in the attempt to find the optimum balance between the level of detail and the possibility to run simulation trajectories for a large number of molecules on very long periods of time, we have chosen a united-atoms representation for the surfactant molecule, as shown in Fig.~\ref{fig:C8}. This is inspired by the model of Allahyarov and Taylor~\cite{allahyarov2008simulation} for the side-chain of Nafion. We represent the hydrophobic section of the molecule with a series of $7$ Lennard-Jones (LJ) neutral beads, each representing an entire \ce{CF2} group. Similarly, the head group is schematized by two charged LJ beads, one for the sulfur atom $S$ and one for the \ce{O3} group, with a total charge $q=-e$. This also allows to associate to the charged head a realistic value for the dipole~\cite{allahyarov2008simulation} and imposes that all acid groups are dissociated. This is a realistic assumption. Indeed, the super-acidity of the sulfonic head groups insures a full ionization of the PFOSA molecule, i.e., hydronium ions are formed at very low water content. This has been proved to be the typical situation for materials containing  sulfur trioxide groups. For instance, recent infra-red kinetics work on the perfluorinated sulfonic acid membrane Nafion~\cite{dalla2014mechanism}, has shown total ionisation already when only one water molecule per sulfonic acid group is present.  

The mass of each bead corresponds to the sum of the atomic masses pertaining to the bead. Details about the beads are given in Table~\ref{table:species}. The topology of the flexible molecules is enforced by harmonic potentials which constrain the distance, the bending angle and the dihedral angle between two, three, and four successive atoms, respectively. The values of the intra-molecular energy parameters used are given in Table~\ref{table:bonded}~\cite{allahyarov2008simulation}. Beads pertaining to different molecules, or located in the same molecule and separated by more than three bonds, interact with the sum of Lennard Jones (LJ) and Coulombic interactions. Numerical values of all re-optimized parameters are given in Table~\ref{table:non-bonded}. Coulombic interactions are truncated and screened according to a modified version of the damped shifted force model~\cite{fennell2006ewald}  which, for a damping parameter $\alpha=0$, simplifies to:
\begin{equation}
V_C(r)=\frac{q_\alpha q_\beta}{4\pi\epsilon_o}
\left[ 
\left( \frac{1}{r}-\frac{1}{r_c}\right)+\frac{1}{r_c^2}(r-r_c)
\right]\theta(r-r_c).
\label{equation:wolf}
\end{equation}
Here $r_c$ is the cut-off, $\epsilon_o=1$ the vacuum dielectric constant, and $\theta$ is the Heavyside function. This potential has been shown to reasonably reproduce the most important features of liquid silica~\cite{carre2007amorphous} and water~\cite{zahn2002enhancement} and is by now considered as an acceptable alternative to methods based on the Ewald summations, both in the bulk and confinement, with an evident gain on the complexity of the calculations. We have set the values of the cutoff to 12\un{\AA}\, and 11\un{\AA}\, for LJ and Colombic interactions, respectively.

Finally, we have considered the three-points rigid SPC/E model for water molecules~\cite{doi:10.1021/j100308a038}, slightly modified to use Eq.~(\ref{equation:wolf}) for the cut-off implementation. We have verified for bulk water that structure, as characterized by total and partial pair distribution functions, and value of the self-diffusion coefficient are in agreement with those generated by the original model. Charge neutrality of the total system is imposed by adding one hydronium ion per acid group, represented by the (four-points) model introduced by Kusaka {\em et al.}~\cite{kusaka_binary_1998}, also modified to use Eq.~(\ref{equation:wolf}). We used the SHAKE algorithm~\cite{allen1987computer} to keep water and hydronium molecules rigid. 

{\bf Simulation details.} We have considered simulation boxes containing $N_S$ surfactants and $N_H=N_S$ hydronium complexes. The number of water molecules $N_W$ was chosen in order to match the needed hydration level. This is quantified by the hydration parameter $\lambda=N_W/N_S$, which counts the number of water molecules per sulfonic acid groups. We considered several values of $\lambda$ in the range $0\div 32$. $\lambda=0$ is the case where hydronium complexes only are present, and therefore does not correspond to the completely dried system. We also considered the case of pure bulk water, as a reference. $N_S$ at each  $\lambda$ has been chosen in order to generate a cubic box edge length of about $100$~\un{\AA} at ambient pressure and temperature conditions. The total number of interacting units is, therefore, $N=N_S(13+3\lambda)$, and varies in the range $80\div 90\times 10^3$ for the considered range of $\lambda$.

We initialized the different systems by placing all molecules in the simulation box, with random positions of the centers of mass and orientations. This allows to avoid introducing any initial preferred order, subsequently letting the system self-organize in the most stable configuration, just based on the interplay among the interaction forces. Initial overlaps were removed by shortly integrating the equations of motion in the $NVE$ ensemble, with the constraint that the displacement of each atom at each time step could not exceed 1~\un{\AA}. Next, we coupled the box to a $P=1$~\un{atm} Berendsen barostat~\cite{allen1987computer} and thermostat, slowly cooling  the system from an initial temperature of $350$\un{K} to the target temperature $T=300$\un{K}. The time step used for the numerical integration of the equation of motions, in this final part of thermalization and for production runs, was $\delta t=1$\un{fs}. Extremely extended thermalisation runs followed, generated according to the $NPT$ ensemble via Nos\'e-Hoover schemes~\cite{allen1987computer}, to allow self-assembly in completely relaxed stable structures.~\footnote{
In the case of the lamellar phases at low hydration levels, the studied time scales were so extended to observe a peculiar form of the `flying ice cube' issue~\cite{Harvey1997}. In this case, the energy of high frequency modes was transferred to the centers of mass of two sub-systems, formed by the absorbed fluid and the surfactant domains, respectively. Next, due to the conservation of {\em total} momentum, the two subsystems moved ballistically in opposite directions, parallel to the planes of the lamellae. We cured this issue by periodically re-sampling the velocities of all atoms from the adequate Maxwell-Boltzmann distribution.} Following the thermalization step, we performed the production runs, dumping complete system configurations (snapshots) at regular time intervals for subsequent analysis. The MD trajectories were integrated by using the high-performance simulation package LAMMPS~\cite{citeulike:2835214}, modified to include the potential in Eq.~(\ref{equation:wolf}).

{\bf Static structure factor.} By using the system configurations extracted from our trajectories, we can calculate the neutron scattering structure factor directly at the microscopic level by
\begin{equation}
S(k) = \left< \frac{N}{\sum_{\alpha}N_{\alpha}b_{\alpha}^{2}} \sum_{\alpha} \sum_{\beta} b_{\alpha} b_{\beta} S_{\alpha\beta}(\vec{k}) \right>_{|\vec{k}|=k},
\label{eq:sk}
\end{equation}
where $b_{\alpha}$ is the coherent neutron scattering length for species $\alpha$ (see Table~\ref{table:species}). In the case of united-atoms beads, $\alpha$ is the sum of the scattering lengths of its components. $\left<\right>$ is the spherical average over wave vectors of modulus $k$. The partial static structure factors involving species $\alpha$ and $\beta$ are defined as
\begin{eqnarray}
S_{\alpha\beta}(\vec{k}) &=& \frac{(1+\delta_{\alpha\beta})}{2N} \rho_{\alpha}(\vec{k})\rho_{\beta}^{*}(\vec{k}), \\
\rho_{\alpha}(\vec{k}) &=& \sum_{l=1}^{N_{\alpha}} \exp\left(i\vec{k}\cdot\vec{r}_l\right),  \nonumber
\label{eq:skab}
\end{eqnarray}
and $\vec{r}_l$ is the instantaneous vector position of bead $l$. In experiments space-averaged information only is accessible, in the form of diffraction diagrams, as those shown in Fig.1 of the main text. These represent as color maps the total scattering intensity, projected on the $(k_x,k_y)-$plane of the detector. Angular integration of these data, at constant $k=\sqrt{k_x^2+k_y^2}$, directly provides $S(k)$ of Eq.~(\ref{eq:sk}).

Information adding to the analysis performed in the main text comes from the partial structure factors, $S_{\alpha\beta}(k)$, shown in Fig.~\ref{fig:skab} for $\lambda=$4 (left) and 32 (right). We show as color-maps all the independent indicated $S_{\alpha\beta}(k)$, with $\alpha\le\beta$ and $\alpha,\beta=\ce{CF_2}, \ce{S},\ce{O_3}, \ce{O_W}, \ce{H_W}, \ce{O_H}, \ce{H_H}$. There are 28 terms in total and we are obviously not interested in scrutinizing all these data in details. We stress a few features, by visual inspection. First, for $\lambda=4$, non-zero signal is concentrated at $k^*\simeq 0.23 \AA^{-1}$ in regions related to correlations of beads pertaining to surfactant or absorbed fluid domains, {\em separately}. There are no significant hydrophobic/aqueous domains correlations contributing constructively to the total signal, and regions of highest intensity correspond to $\ce{CF_2}-\ce{CF_2}$, $\ce{O_W}-\ce{O_W}$, $\ce{O_W}-\ce{H_H}$ and $\ce{H_W}-\ce{H_W}$ correlations. In contrast, a certain degree of anti-correlation, producing negative values for $S_{\alpha\beta}(k)$ at the same $k^*$, is associated to interactions of the $\ce{CF_2}$ beads with absorbed water atoms. The intensity of the diffraction peak therefore mainly comes from independent contributions due to the confining matrix and the absorbed fluids, with a (mild) destructive interference associated to the interfaces. The fact that the anti-correlated negative signal is also localized at $k^*$ is not surprising, if we note that the associated length scale $l^*=2\pi/k^*\simeq 27\AA$ is related to the repetition period of like-domains. This implies that starting from the surfactant (ionic) phase and moving of a distance $l^*$, a system's unit will never reach the ionic (surfactant) phase. In the case of $\lambda=32$ (right) the situation is in general qualitatively similar, with the maximum intensity region shifted to lower values of $k$, and the total constructive signal being spread on a larger $k-$range, due to disorder. However, the signal associated to both hydronium/hydronium and hydronium/water correlations and due to the ionic condensation at the interface at very low hydration is completely suppressed, following significant solvation of ions, as we show in the main text.
\bibliography{bibliography}

\begin{thebibliography}{48}%
\makeatletter
\providecommand \@ifxundefined [1]{%
 \@ifx{#1\undefined}
}%
\providecommand \@ifnum [1]{%
 \ifnum #1\expandafter \@firstoftwo
 \else \expandafter \@secondoftwo
 \fi
}%
\providecommand \@ifx [1]{%
 \ifx #1\expandafter \@firstoftwo
 \else \expandafter \@secondoftwo
 \fi
}%
\providecommand \natexlab [1]{#1}%
\providecommand \enquote  [1]{``#1''}%
\providecommand \bibnamefont  [1]{#1}%
\providecommand \bibfnamefont [1]{#1}%
\providecommand \citenamefont [1]{#1}%
\providecommand \href@noop [0]{\@secondoftwo}%
\providecommand \href [0]{\begingroup \@sanitize@url \@href}%
\providecommand \@href[1]{\@@startlink{#1}\@@href}%
\providecommand \@@href[1]{\endgroup#1\@@endlink}%
\providecommand \@sanitize@url [0]{\catcode `\\12\catcode `\$12\catcode
  `\&12\catcode `\#12\catcode `\^12\catcode `\_12\catcode `\%12\relax}%
\providecommand \@@startlink[1]{}%
\providecommand \@@endlink[0]{}%
\providecommand \url  [0]{\begingroup\@sanitize@url \@url }%
\providecommand \@url [1]{\endgroup\@href {#1}{\urlprefix }}%
\providecommand \urlprefix  [0]{URL }%
\providecommand \Eprint [0]{\href }%
\providecommand \doibase [0]{http://dx.doi.org/}%
\providecommand \selectlanguage [0]{\@gobble}%
\providecommand \bibinfo  [0]{\@secondoftwo}%
\providecommand \bibfield  [0]{\@secondoftwo}%
\providecommand \translation [1]{[#1]}%
\providecommand \BibitemOpen [0]{}%
\providecommand \bibitemStop [0]{}%
\providecommand \bibitemNoStop [0]{.\EOS\space}%
\providecommand \EOS [0]{\spacefactor3000\relax}%
\providecommand \BibitemShut  [1]{\csname bibitem#1\endcsname}%
\let\auto@bib@innerbib\@empty
\bibitem [{\citenamefont {Jones}(2002)}]{jones2002soft}%
  \BibitemOpen
  \bibfield  {author} {\bibinfo {author} {\bibfnamefont {R.~A.}\ \bibnamefont
  {Jones}},\ }\href@noop {} {\emph {\bibinfo {title} {Soft condensed matter}}}\
  (\bibinfo  {publisher} {IOP Publishing},\ \bibinfo {year} {2002})\BibitemShut
  {NoStop}%
\bibitem [{\citenamefont {Schramm}\ \emph {et~al.}(2003)\citenamefont
  {Schramm}, \citenamefont {Stasiuk},\ and\ \citenamefont
  {Marangoni}}]{schramm20032}%
  \BibitemOpen
  \bibfield  {author} {\bibinfo {author} {\bibfnamefont {L.~L.}\ \bibnamefont
  {Schramm}}, \bibinfo {author} {\bibfnamefont {E.~N.}\ \bibnamefont
  {Stasiuk}}, \ and\ \bibinfo {author} {\bibfnamefont {D.~G.}\ \bibnamefont
  {Marangoni}},\ }\href@noop {} {\bibfield  {journal} {\bibinfo  {journal}
  {Annual Reports Section" C"(Physical Chemistry)}\ }\textbf {\bibinfo {volume}
  {99}},\ \bibinfo {pages} {3} (\bibinfo {year} {2003})}\BibitemShut {NoStop}%
\bibitem [{\citenamefont {Boek}\ \emph {et~al.}(2002)\citenamefont {Boek},
  \citenamefont {Jusufi}, \citenamefont {L{\"o}wen},\ and\ \citenamefont
  {Maitland}}]{boek2002molecular}%
  \BibitemOpen
  \bibfield  {author} {\bibinfo {author} {\bibfnamefont {E.}~\bibnamefont
  {Boek}}, \bibinfo {author} {\bibfnamefont {A.}~\bibnamefont {Jusufi}},
  \bibinfo {author} {\bibfnamefont {H.}~\bibnamefont {L{\"o}wen}}, \ and\
  \bibinfo {author} {\bibfnamefont {G.}~\bibnamefont {Maitland}},\ }\href@noop
  {} {\bibfield  {journal} {\bibinfo  {journal} {Journal of Physics: Condensed
  Matter}\ }\textbf {\bibinfo {volume} {14}},\ \bibinfo {pages} {9413}
  (\bibinfo {year} {2002})}\BibitemShut {NoStop}%
\bibitem [{\citenamefont {Faul}\ and\ \citenamefont
  {Antonietti}(2003)}]{faul2003ionic}%
  \BibitemOpen
  \bibfield  {author} {\bibinfo {author} {\bibfnamefont {C.~F.}\ \bibnamefont
  {Faul}}\ and\ \bibinfo {author} {\bibfnamefont {M.}~\bibnamefont
  {Antonietti}},\ }\href@noop {} {\bibfield  {journal} {\bibinfo  {journal}
  {Advanced Materials}\ }\textbf {\bibinfo {volume} {15}},\ \bibinfo {pages}
  {673} (\bibinfo {year} {2003})}\BibitemShut {NoStop}%
\bibitem [{\citenamefont {Zemb}\ and\ \citenamefont
  {Lindner}(2002)}]{zemb2002neutrons}%
  \BibitemOpen
  \bibfield  {author} {\bibinfo {author} {\bibfnamefont {T.}~\bibnamefont
  {Zemb}}\ and\ \bibinfo {author} {\bibfnamefont {P.}~\bibnamefont {Lindner}},\
  }\href@noop {} {\emph {\bibinfo {title} {Neutrons, X-rays and light:
  scattering methods applied to soft condensed matter}}}\ (\bibinfo
  {publisher} {North Holland},\ \bibinfo {year} {2002})\BibitemShut {NoStop}%
\bibitem [{\citenamefont {Shelley}\ and\ \citenamefont
  {Shelley}(2000)}]{shelley2000computer}%
  \BibitemOpen
  \bibfield  {author} {\bibinfo {author} {\bibfnamefont {J.~C.}\ \bibnamefont
  {Shelley}}\ and\ \bibinfo {author} {\bibfnamefont {M.~Y.}\ \bibnamefont
  {Shelley}},\ }\href@noop {} {\bibfield  {journal} {\bibinfo  {journal}
  {Current opinion in colloid \& interface science}\ }\textbf {\bibinfo
  {volume} {5}},\ \bibinfo {pages} {101} (\bibinfo {year} {2000})}\BibitemShut
  {NoStop}%
\bibitem [{\citenamefont {Alcoutlabi}\ and\ \citenamefont
  {McKenna}(2005)}]{alcoutlabi2005effects}%
  \BibitemOpen
  \bibfield  {author} {\bibinfo {author} {\bibfnamefont {M.}~\bibnamefont
  {Alcoutlabi}}\ and\ \bibinfo {author} {\bibfnamefont {G.~B.}\ \bibnamefont
  {McKenna}},\ }\href@noop {} {\bibfield  {journal} {\bibinfo  {journal}
  {Journal of Physics: Condensed Matter}\ }\textbf {\bibinfo {volume} {17}},\
  \bibinfo {pages} {R461} (\bibinfo {year} {2005})}\BibitemShut {NoStop}%
\bibitem [{\citenamefont {Frick}\ \emph {et~al.}(2005)\citenamefont {Frick},
  \citenamefont {Alba-Simionesco}, \citenamefont {Dosseh}, \citenamefont
  {Le~Quellec}, \citenamefont {Moreno}, \citenamefont {Colmenero},
  \citenamefont {Sch{\"o}nhals}, \citenamefont {Zorn}, \citenamefont
  {Chrissopoulou}, \citenamefont {Anastasiadis},\ and\ \citenamefont
  {Dalnoki-Veress}}]{frick2005inelastic}%
  \BibitemOpen
  \bibfield  {author} {\bibinfo {author} {\bibfnamefont {B.}~\bibnamefont
  {Frick}}, \bibinfo {author} {\bibfnamefont {C.}~\bibnamefont
  {Alba-Simionesco}}, \bibinfo {author} {\bibfnamefont {G.}~\bibnamefont
  {Dosseh}}, \bibinfo {author} {\bibfnamefont {C.}~\bibnamefont {Le~Quellec}},
  \bibinfo {author} {\bibfnamefont {A.}~\bibnamefont {Moreno}}, \bibinfo
  {author} {\bibfnamefont {J.}~\bibnamefont {Colmenero}}, \bibinfo {author}
  {\bibfnamefont {A.}~\bibnamefont {Sch{\"o}nhals}}, \bibinfo {author}
  {\bibfnamefont {R.}~\bibnamefont {Zorn}}, \bibinfo {author} {\bibfnamefont
  {K.}~\bibnamefont {Chrissopoulou}}, \bibinfo {author} {\bibfnamefont
  {S.}~\bibnamefont {Anastasiadis}}, \ and\ \bibinfo {author} {\bibfnamefont
  {K.}~\bibnamefont {Dalnoki-Veress}},\ }\href@noop {} {\bibfield  {journal}
  {\bibinfo  {journal} {Journal of non-crystalline solids}\ }\textbf {\bibinfo
  {volume} {351}},\ \bibinfo {pages} {2657} (\bibinfo {year}
  {2005})}\BibitemShut {NoStop}%
\bibitem [{\citenamefont {Alba-Simionesco}\ \emph {et~al.}(2006)\citenamefont
  {Alba-Simionesco}, \citenamefont {Coasne}, \citenamefont {Dosseh},
  \citenamefont {Dudziak}, \citenamefont {Gubbins}, \citenamefont
  {Radhakrishnan},\ and\ \citenamefont
  {Sliwinska-Bartkowiak}}]{alba2006effects}%
  \BibitemOpen
  \bibfield  {author} {\bibinfo {author} {\bibfnamefont {C.}~\bibnamefont
  {Alba-Simionesco}}, \bibinfo {author} {\bibfnamefont {B.}~\bibnamefont
  {Coasne}}, \bibinfo {author} {\bibfnamefont {G.}~\bibnamefont {Dosseh}},
  \bibinfo {author} {\bibfnamefont {G.}~\bibnamefont {Dudziak}}, \bibinfo
  {author} {\bibfnamefont {K.}~\bibnamefont {Gubbins}}, \bibinfo {author}
  {\bibfnamefont {R.}~\bibnamefont {Radhakrishnan}}, \ and\ \bibinfo {author}
  {\bibfnamefont {M.}~\bibnamefont {Sliwinska-Bartkowiak}},\ }\href@noop {}
  {\bibfield  {journal} {\bibinfo  {journal} {Journal of Physics: Condensed
  Matter}\ }\textbf {\bibinfo {volume} {18}},\ \bibinfo {pages} {R15} (\bibinfo
  {year} {2006})}\BibitemShut {NoStop}%
\bibitem [{\citenamefont {Rasaiah}\ \emph {et~al.}(2008)\citenamefont
  {Rasaiah}, \citenamefont {Garde},\ and\ \citenamefont
  {Hummer}}]{rasaiah2008water}%
  \BibitemOpen
  \bibfield  {author} {\bibinfo {author} {\bibfnamefont {J.~C.}\ \bibnamefont
  {Rasaiah}}, \bibinfo {author} {\bibfnamefont {S.}~\bibnamefont {Garde}}, \
  and\ \bibinfo {author} {\bibfnamefont {G.}~\bibnamefont {Hummer}},\
  }\href@noop {} {\bibfield  {journal} {\bibinfo  {journal} {Annu. Rev. Phys.
  Chem.}\ }\textbf {\bibinfo {volume} {59}},\ \bibinfo {pages} {713} (\bibinfo
  {year} {2008})}\BibitemShut {NoStop}%
\bibitem [{\citenamefont {Zanotti}\ \emph {et~al.}(2012)\citenamefont
  {Zanotti}, \citenamefont {Lagren{\'e}}, \citenamefont {Malikova},
  \citenamefont {Judeinstein}, \citenamefont {Panesar}, \citenamefont
  {Ollivier}, \citenamefont {Rols}, \citenamefont {Mayne-L’Hermite},
  \citenamefont {Pinault},\ and\ \citenamefont
  {Boulanger}}]{zanotti2012nanometric}%
  \BibitemOpen
  \bibfield  {author} {\bibinfo {author} {\bibfnamefont {J.-M.}\ \bibnamefont
  {Zanotti}}, \bibinfo {author} {\bibfnamefont {K.}~\bibnamefont
  {Lagren{\'e}}}, \bibinfo {author} {\bibfnamefont {N.}~\bibnamefont
  {Malikova}}, \bibinfo {author} {\bibfnamefont {P.}~\bibnamefont
  {Judeinstein}}, \bibinfo {author} {\bibfnamefont {K.}~\bibnamefont
  {Panesar}}, \bibinfo {author} {\bibfnamefont {J.}~\bibnamefont {Ollivier}},
  \bibinfo {author} {\bibfnamefont {S.}~\bibnamefont {Rols}}, \bibinfo {author}
  {\bibfnamefont {M.}~\bibnamefont {Mayne-L’Hermite}}, \bibinfo {author}
  {\bibfnamefont {M.}~\bibnamefont {Pinault}}, \ and\ \bibinfo {author}
  {\bibfnamefont {P.}~\bibnamefont {Boulanger}},\ }\href@noop {} {\bibfield
  {journal} {\bibinfo  {journal} {The European Physical Journal Special
  Topics}\ }\textbf {\bibinfo {volume} {213}},\ \bibinfo {pages} {129}
  (\bibinfo {year} {2012})}\BibitemShut {NoStop}%
\bibitem [{\citenamefont {Perkin}\ and\ \citenamefont
  {Klein}(2013)}]{perkin2013soft}%
  \BibitemOpen
  \bibfield  {author} {\bibinfo {author} {\bibfnamefont {S.}~\bibnamefont
  {Perkin}}\ and\ \bibinfo {author} {\bibfnamefont {J.}~\bibnamefont {Klein}},\
  }\href {\doibase 10.1039/C3SM90141F} {\bibfield  {journal} {\bibinfo
  {journal} {Soft Matter}\ }\textbf {\bibinfo {volume} {9}},\ \bibinfo {pages}
  {10438} (\bibinfo {year} {2013})}\BibitemShut {NoStop}%
\bibitem [{\citenamefont {Wang}\ \emph {et~al.}(2004)\citenamefont {Wang},
  \citenamefont {He},\ and\ \citenamefont {Richert}}]{wang2004intramicellar}%
  \BibitemOpen
  \bibfield  {author} {\bibinfo {author} {\bibfnamefont {L.-M.}\ \bibnamefont
  {Wang}}, \bibinfo {author} {\bibfnamefont {F.}~\bibnamefont {He}}, \ and\
  \bibinfo {author} {\bibfnamefont {R.}~\bibnamefont {Richert}},\ }\href@noop
  {} {\bibfield  {journal} {\bibinfo  {journal} {Physical review letters}\
  }\textbf {\bibinfo {volume} {92}},\ \bibinfo {pages} {095701} (\bibinfo
  {year} {2004})}\BibitemShut {NoStop}%
\bibitem [{\citenamefont {Hunter}\ \emph {et~al.}(2014)\citenamefont {Hunter},
  \citenamefont {Edmond},\ and\ \citenamefont {Weeks}}]{hunter2014boundary}%
  \BibitemOpen
  \bibfield  {author} {\bibinfo {author} {\bibfnamefont {G.~L.}\ \bibnamefont
  {Hunter}}, \bibinfo {author} {\bibfnamefont {K.~V.}\ \bibnamefont {Edmond}},
  \ and\ \bibinfo {author} {\bibfnamefont {E.~R.}\ \bibnamefont {Weeks}},\
  }\href@noop {} {\bibfield  {journal} {\bibinfo  {journal} {Physical Review
  Letters}\ }\textbf {\bibinfo {volume} {112}},\ \bibinfo {pages} {218302}
  (\bibinfo {year} {2014})}\BibitemShut {NoStop}%
\bibitem [{\citenamefont {Mauritz}\ and\ \citenamefont
  {Moore}(2004)}]{mauritz2004state}%
  \BibitemOpen
  \bibfield  {author} {\bibinfo {author} {\bibfnamefont {K.~A.}\ \bibnamefont
  {Mauritz}}\ and\ \bibinfo {author} {\bibfnamefont {R.~B.}\ \bibnamefont
  {Moore}},\ }\href@noop {} {\bibfield  {journal} {\bibinfo  {journal}
  {Chemical reviews}\ }\textbf {\bibinfo {volume} {104}},\ \bibinfo {pages}
  {4535} (\bibinfo {year} {2004})}\BibitemShut {NoStop}%
\bibitem [{\citenamefont {Lyonnard}\ \emph {et~al.}(2010)\citenamefont
  {Lyonnard}, \citenamefont {Berrod}, \citenamefont {Br{\"u}ning},
  \citenamefont {Gebel}, \citenamefont {Guillermo}, \citenamefont {Ftouni},
  \citenamefont {Ollivier},\ and\ \citenamefont
  {Frick}}]{lyonnard2010perfluorinated}%
  \BibitemOpen
  \bibfield  {author} {\bibinfo {author} {\bibfnamefont {S.}~\bibnamefont
  {Lyonnard}}, \bibinfo {author} {\bibfnamefont {Q.}~\bibnamefont {Berrod}},
  \bibinfo {author} {\bibfnamefont {B.-A.}\ \bibnamefont {Br{\"u}ning}},
  \bibinfo {author} {\bibfnamefont {G.}~\bibnamefont {Gebel}}, \bibinfo
  {author} {\bibfnamefont {A.}~\bibnamefont {Guillermo}}, \bibinfo {author}
  {\bibfnamefont {H.}~\bibnamefont {Ftouni}}, \bibinfo {author} {\bibfnamefont
  {J.}~\bibnamefont {Ollivier}}, \ and\ \bibinfo {author} {\bibfnamefont
  {B.}~\bibnamefont {Frick}},\ }\href@noop {} {\bibfield  {journal} {\bibinfo
  {journal} {The European Physical Journal-Special Topics}\ }\textbf {\bibinfo
  {volume} {189}},\ \bibinfo {pages} {205} (\bibinfo {year}
  {2010})}\BibitemShut {NoStop}%
\bibitem [{\citenamefont {Berrod}\ \emph {et~al.}(2014)\citenamefont {Berrod},
  \citenamefont {Lyonnard}, \citenamefont {Guillermo}, \citenamefont
  {Ollivier}, \citenamefont {Frick},\ and\ \citenamefont {Gebel}}]{berrod2014}%
  \BibitemOpen
  \bibfield  {author} {\bibinfo {author} {\bibfnamefont {Q.}~\bibnamefont
  {Berrod}}, \bibinfo {author} {\bibfnamefont {S.}~\bibnamefont {Lyonnard}},
  \bibinfo {author} {\bibfnamefont {A.}~\bibnamefont {Guillermo}}, \bibinfo
  {author} {\bibfnamefont {J.}~\bibnamefont {Ollivier}}, \bibinfo {author}
  {\bibfnamefont {B.}~\bibnamefont {Frick}}, \ and\ \bibinfo {author}
  {\bibfnamefont {G.}~\bibnamefont {Gebel}},\ }\href@noop {} {\bibfield
  {journal} {\bibinfo  {journal} {The European Physical Journal}\ ,\ \bibinfo
  {pages} {in press}} (\bibinfo {year} {2014})}\BibitemShut {NoStop}%
\bibitem [{\citenamefont {Kissa}(2001)}]{kissa2001fluorinated}%
  \BibitemOpen
  \bibfield  {author} {\bibinfo {author} {\bibfnamefont {E.}~\bibnamefont
  {Kissa}},\ }\href@noop {} {\emph {\bibinfo {title} {Fluorinated surfactants
  and repellents}}}\ (\bibinfo  {publisher} {CRC Press},\ \bibinfo {year}
  {2001})\BibitemShut {NoStop}%
\bibitem [{\citenamefont {Allahyarov}\ and\ \citenamefont
  {Taylor}(2008)}]{allahyarov2008simulation}%
  \BibitemOpen
  \bibfield  {author} {\bibinfo {author} {\bibfnamefont {E.}~\bibnamefont
  {Allahyarov}}\ and\ \bibinfo {author} {\bibfnamefont {P.~L.}\ \bibnamefont
  {Taylor}},\ }\href@noop {} {\bibfield  {journal} {\bibinfo  {journal} {The
  Journal of Physical Chemistry B}\ }\textbf {\bibinfo {volume} {113}},\
  \bibinfo {pages} {610} (\bibinfo {year} {2008})}\BibitemShut {NoStop}%
\bibitem [{\citenamefont {Kreuer}(2013)}]{kreuer2013ion}%
  \BibitemOpen
  \bibfield  {author} {\bibinfo {author} {\bibfnamefont {K.-D.}\ \bibnamefont
  {Kreuer}},\ }\href@noop {} {\bibfield  {journal} {\bibinfo  {journal}
  {Chemistry of Materials}\ }\textbf {\bibinfo {volume} {26}},\ \bibinfo
  {pages} {361} (\bibinfo {year} {2013})}\BibitemShut {NoStop}%
\bibitem [{\citenamefont {Gebel}\ and\ \citenamefont
  {Diat}(2005)}]{gebel2005neutron}%
  \BibitemOpen
  \bibfield  {author} {\bibinfo {author} {\bibfnamefont {G.}~\bibnamefont
  {Gebel}}\ and\ \bibinfo {author} {\bibfnamefont {O.}~\bibnamefont {Diat}},\
  }\href@noop {} {\bibfield  {journal} {\bibinfo  {journal} {Fuel Cells}\
  }\textbf {\bibinfo {volume} {5}},\ \bibinfo {pages} {261} (\bibinfo {year}
  {2005})}\BibitemShut {NoStop}%
\bibitem [{Note1()}]{Note1}%
  \BibitemOpen
  \bibinfo {note} {We note that Ref.~\cite {sunda2013molecular} contains an
  extremely detailed all-atoms simulation study of the side chain pendants of
  three perfluorosulfonic acid polymer electrolyte membranes, characterized by
  different structures. Structural organization was investigated in details,
  together with some transport properties, at different degrees of hydration.
  Interestingly, the Authors completely overlooked the possibility to comment
  on similarities with ionic surfactant phases.}\BibitemShut {Stop}%
\bibitem [{\citenamefont {Sood}\ \emph {et~al.}(2012)\citenamefont {Sood},
  \citenamefont {Iojoiu}, \citenamefont {Espuche}, \citenamefont {Gouanvé},
  \citenamefont {Gebel}, \citenamefont {Mendil-Jakani}, \citenamefont
  {Lyonnard},\ and\ \citenamefont {Jestin}}]{sood2012proton}%
  \BibitemOpen
  \bibfield  {author} {\bibinfo {author} {\bibfnamefont {R.}~\bibnamefont
  {Sood}}, \bibinfo {author} {\bibfnamefont {C.}~\bibnamefont {Iojoiu}},
  \bibinfo {author} {\bibfnamefont {E.}~\bibnamefont {Espuche}}, \bibinfo
  {author} {\bibfnamefont {F.}~\bibnamefont {Gouanvé}}, \bibinfo {author}
  {\bibfnamefont {G.}~\bibnamefont {Gebel}}, \bibinfo {author} {\bibfnamefont
  {H.}~\bibnamefont {Mendil-Jakani}}, \bibinfo {author} {\bibfnamefont
  {S.}~\bibnamefont {Lyonnard}}, \ and\ \bibinfo {author} {\bibfnamefont
  {J.}~\bibnamefont {Jestin}},\ }\href@noop {} {\bibfield  {journal} {\bibinfo
  {journal} {The Journal of Physical Chemistry C}\ }\textbf {\bibinfo {volume}
  {116}},\ \bibinfo {pages} {24413} (\bibinfo {year} {2012})}\BibitemShut
  {NoStop}%
\bibitem [{\citenamefont {Lu}\ \emph {et~al.}(2014)\citenamefont {Lu},
  \citenamefont {Gao}, \citenamefont {Xie}, \citenamefont {Sun},\ and\
  \citenamefont {Zheng}}]{lu2014chemical}%
  \BibitemOpen
  \bibfield  {author} {\bibinfo {author} {\bibfnamefont {F.}~\bibnamefont
  {Lu}}, \bibinfo {author} {\bibfnamefont {X.}~\bibnamefont {Gao}}, \bibinfo
  {author} {\bibfnamefont {S.}~\bibnamefont {Xie}}, \bibinfo {author}
  {\bibfnamefont {N.}~\bibnamefont {Sun}}, \ and\ \bibinfo {author}
  {\bibfnamefont {L.}~\bibnamefont {Zheng}},\ }\href@noop {} {\bibfield
  {journal} {\bibinfo  {journal} {Soft matter}\ }\textbf {\bibinfo {volume}
  {10}},\ \bibinfo {pages} {7819} (\bibinfo {year} {2014})}\BibitemShut
  {NoStop}%
\bibitem [{\citenamefont {Chandler}(2005)}]{chandler2005interfaces}%
  \BibitemOpen
  \bibfield  {author} {\bibinfo {author} {\bibfnamefont {D.}~\bibnamefont
  {Chandler}},\ }\href@noop {} {\bibfield  {journal} {\bibinfo  {journal}
  {Nature}\ }\textbf {\bibinfo {volume} {437}},\ \bibinfo {pages} {640}
  (\bibinfo {year} {2005})}\BibitemShut {NoStop}%
\bibitem [{\citenamefont {Willard}\ and\ \citenamefont
  {Chandler}(2014)}]{willard2014molecular}%
  \BibitemOpen
  \bibfield  {author} {\bibinfo {author} {\bibfnamefont {A.~P.}\ \bibnamefont
  {Willard}}\ and\ \bibinfo {author} {\bibfnamefont {D.}~\bibnamefont
  {Chandler}},\ }\href@noop {} {\bibfield  {journal} {\bibinfo  {journal}
  {arXiv:1407.4365 [cond-mat.soft]}\ } (\bibinfo {year} {2014})}\BibitemShut
  {NoStop}%
\bibitem [{\citenamefont {Chandler}\ and\ \citenamefont
  {Varilly}(2011)}]{chandler2011lectures}%
  \BibitemOpen
  \bibfield  {author} {\bibinfo {author} {\bibfnamefont {D.}~\bibnamefont
  {Chandler}}\ and\ \bibinfo {author} {\bibfnamefont {P.}~\bibnamefont
  {Varilly}},\ }\href@noop {} {\bibfield  {journal} {\bibinfo  {journal} {arXiv
  preprint arXiv:1101.2235}\ } (\bibinfo {year} {2011})}\BibitemShut {NoStop}%
\bibitem [{\citenamefont {Bizzarri}\ and\ \citenamefont
  {Cannistraro}(2002)}]{bizzarri2002molecular}%
  \BibitemOpen
  \bibfield  {author} {\bibinfo {author} {\bibfnamefont {A.~R.}\ \bibnamefont
  {Bizzarri}}\ and\ \bibinfo {author} {\bibfnamefont {S.}~\bibnamefont
  {Cannistraro}},\ }\href@noop {} {\bibfield  {journal} {\bibinfo  {journal}
  {The Journal of Physical Chemistry B}\ }\textbf {\bibinfo {volume} {106}},\
  \bibinfo {pages} {6617} (\bibinfo {year} {2002})}\BibitemShut {NoStop}%
\bibitem [{\citenamefont {Yamamoto}\ \emph {et~al.}(2014)\citenamefont
  {Yamamoto}, \citenamefont {Akimoto}, \citenamefont {Yasui},\ and\
  \citenamefont {Yasuoka}}]{yamamoto2014origin}%
  \BibitemOpen
  \bibfield  {author} {\bibinfo {author} {\bibfnamefont {E.}~\bibnamefont
  {Yamamoto}}, \bibinfo {author} {\bibfnamefont {T.}~\bibnamefont {Akimoto}},
  \bibinfo {author} {\bibfnamefont {M.}~\bibnamefont {Yasui}}, \ and\ \bibinfo
  {author} {\bibfnamefont {K.}~\bibnamefont {Yasuoka}},\ }\href@noop {}
  {\bibfield  {journal} {\bibinfo  {journal} {Scientific reports}\ }\textbf
  {\bibinfo {volume} {4}} (\bibinfo {year} {2014})}\BibitemShut {NoStop}%
\bibitem [{\citenamefont {Jorge}(2008)}]{Jorge2008}%
  \BibitemOpen
  \bibfield  {author} {\bibinfo {author} {\bibfnamefont {M.}~\bibnamefont
  {Jorge}},\ }\href {\doibase 10.1021/la800291p} {\bibfield  {journal}
  {\bibinfo  {journal} {Langmuir : the ACS journal of surfaces and colloids}\
  }\textbf {\bibinfo {volume} {24}},\ \bibinfo {pages} {5714} (\bibinfo {year}
  {2008})}\BibitemShut {NoStop}%
\bibitem [{\citenamefont {Jusufi}\ \emph {et~al.}(2009)\citenamefont {Jusufi},
  \citenamefont {Hynninen}, \citenamefont {Haataja},\ and\ \citenamefont
  {Panagiotopoulos}}]{Jusufi2009}%
  \BibitemOpen
  \bibfield  {author} {\bibinfo {author} {\bibfnamefont {A.}~\bibnamefont
  {Jusufi}}, \bibinfo {author} {\bibfnamefont {A.-P.}\ \bibnamefont
  {Hynninen}}, \bibinfo {author} {\bibfnamefont {M.}~\bibnamefont {Haataja}}, \
  and\ \bibinfo {author} {\bibfnamefont {A.~Z.}\ \bibnamefont
  {Panagiotopoulos}},\ }\href {\doibase 10.1021/jp901032g} {\bibfield
  {journal} {\bibinfo  {journal} {The journal of physical chemistry. B}\
  }\textbf {\bibinfo {volume} {113}},\ \bibinfo {pages} {6314} (\bibinfo {year}
  {2009})}\BibitemShut {NoStop}%
\bibitem [{\citenamefont {Sayyed-Ahmad}\ \emph {et~al.}(2010)\citenamefont
  {Sayyed-Ahmad}, \citenamefont {Lichtenberger},\ and\ \citenamefont
  {Gorfe}}]{Sayyed-Ahmad2010}%
  \BibitemOpen
  \bibfield  {author} {\bibinfo {author} {\bibfnamefont {A.}~\bibnamefont
  {Sayyed-Ahmad}}, \bibinfo {author} {\bibfnamefont {L.~M.}\ \bibnamefont
  {Lichtenberger}}, \ and\ \bibinfo {author} {\bibfnamefont {A.~A.}\
  \bibnamefont {Gorfe}},\ }\href {\doibase 10.1021/la102106t} {\bibfield
  {journal} {\bibinfo  {journal} {Langmuir : the ACS journal of surfaces and
  colloids}\ }\textbf {\bibinfo {volume} {26}},\ \bibinfo {pages} {13407}
  (\bibinfo {year} {2010})}\BibitemShut {NoStop}%
\bibitem [{\citenamefont {Sanders}\ and\ \citenamefont
  {Panagiotopoulos}(2010)}]{Sanders2010}%
  \BibitemOpen
  \bibfield  {author} {\bibinfo {author} {\bibfnamefont {S.~A.}\ \bibnamefont
  {Sanders}}\ and\ \bibinfo {author} {\bibfnamefont {A.~Z.}\ \bibnamefont
  {Panagiotopoulos}},\ }\href {\doibase 10.1063/1.3358354} {\bibfield
  {journal} {\bibinfo  {journal} {The Journal of chemical physics}\ }\textbf
  {\bibinfo {volume} {132}},\ \bibinfo {pages} {114902} (\bibinfo {year}
  {2010})}\BibitemShut {NoStop}%
\bibitem [{\citenamefont {Shinoda}\ \emph
  {et~al.}(2008{\natexlab{a}})\citenamefont {Shinoda}, \citenamefont {DeVane},\
  and\ \citenamefont {Klein}}]{Shinoda2008}%
  \BibitemOpen
  \bibfield  {author} {\bibinfo {author} {\bibfnamefont {W.}~\bibnamefont
  {Shinoda}}, \bibinfo {author} {\bibfnamefont {R.}~\bibnamefont {DeVane}}, \
  and\ \bibinfo {author} {\bibfnamefont {M.~L.}\ \bibnamefont {Klein}},\ }\href
  {\doibase 10.1039/b808701f} {\bibfield  {journal} {\bibinfo  {journal} {Soft
  Matter}\ }\textbf {\bibinfo {volume} {4}},\ \bibinfo {pages} {2454} (\bibinfo
  {year} {2008}{\natexlab{a}})}\BibitemShut {NoStop}%
\bibitem [{\citenamefont {Klein}\ and\ \citenamefont
  {Shinoda}(2008)}]{Klein2008}%
  \BibitemOpen
  \bibfield  {author} {\bibinfo {author} {\bibfnamefont {M.~L.}\ \bibnamefont
  {Klein}}\ and\ \bibinfo {author} {\bibfnamefont {W.}~\bibnamefont
  {Shinoda}},\ }\href {\doibase 10.1126/science.1157834} {\bibfield  {journal}
  {\bibinfo  {journal} {Science (New York, N.Y.)}\ }\textbf {\bibinfo {volume}
  {321}},\ \bibinfo {pages} {798} (\bibinfo {year} {2008})}\BibitemShut
  {NoStop}%
\bibitem [{\citenamefont {Bhattacharya}\ and\ \citenamefont
  {Mahanti}(2001)}]{bhattacharya2001self}%
  \BibitemOpen
  \bibfield  {author} {\bibinfo {author} {\bibfnamefont {A.}~\bibnamefont
  {Bhattacharya}}\ and\ \bibinfo {author} {\bibfnamefont {S.}~\bibnamefont
  {Mahanti}},\ }\href@noop {} {\bibfield  {journal} {\bibinfo  {journal}
  {Journal of Physics: Condensed Matter}\ }\textbf {\bibinfo {volume} {13}},\
  \bibinfo {pages} {1413} (\bibinfo {year} {2001})}\BibitemShut {NoStop}%
\bibitem [{\citenamefont {Shinoda}\ \emph
  {et~al.}(2008{\natexlab{b}})\citenamefont {Shinoda}, \citenamefont {DeVane},\
  and\ \citenamefont {Klein}}]{Shinoda2008a}%
  \BibitemOpen
  \bibfield  {author} {\bibinfo {author} {\bibfnamefont {W.}~\bibnamefont
  {Shinoda}}, \bibinfo {author} {\bibfnamefont {R.}~\bibnamefont {DeVane}}, \
  and\ \bibinfo {author} {\bibfnamefont {M.~L.}\ \bibnamefont {Klein}},\ }\href
  {\doibase 10.1039/b808701f} {\bibfield  {journal} {\bibinfo  {journal} {Soft
  Matter}\ }\textbf {\bibinfo {volume} {4}},\ \bibinfo {pages} {2454} (\bibinfo
  {year} {2008}{\natexlab{b}})}\BibitemShut {NoStop}%
\bibitem [{\citenamefont {Dalla~Bernardina}\ \emph {et~al.}(2014)\citenamefont
  {Dalla~Bernardina}, \citenamefont {Brubach}, \citenamefont {Berrod},
  \citenamefont {Guillermo}, \citenamefont {Judeinstein}, \citenamefont {Roy},\
  and\ \citenamefont {Lyonnard}}]{dalla2014mechanism}%
  \BibitemOpen
  \bibfield  {author} {\bibinfo {author} {\bibfnamefont {S.}~\bibnamefont
  {Dalla~Bernardina}}, \bibinfo {author} {\bibfnamefont {J.-B.}\ \bibnamefont
  {Brubach}}, \bibinfo {author} {\bibfnamefont {Q.}~\bibnamefont {Berrod}},
  \bibinfo {author} {\bibfnamefont {A.}~\bibnamefont {Guillermo}}, \bibinfo
  {author} {\bibfnamefont {P.}~\bibnamefont {Judeinstein}}, \bibinfo {author}
  {\bibfnamefont {P.}~\bibnamefont {Roy}}, \ and\ \bibinfo {author}
  {\bibfnamefont {S.}~\bibnamefont {Lyonnard}},\ }\href {\doibase
  10.1021/jp5074818} {\bibfield  {journal} {\bibinfo  {journal} {The Journal of
  Physical Chemistry C}\ } (\bibinfo {year} {2014}),\
  10.1021/jp5074818}\BibitemShut {NoStop}%
\bibitem [{\citenamefont {Fennell}\ and\ \citenamefont
  {Gezelter}(2006)}]{fennell2006ewald}%
  \BibitemOpen
  \bibfield  {author} {\bibinfo {author} {\bibfnamefont {C.~J.}\ \bibnamefont
  {Fennell}}\ and\ \bibinfo {author} {\bibfnamefont {J.~D.}\ \bibnamefont
  {Gezelter}},\ }\href@noop {} {\bibfield  {journal} {\bibinfo  {journal} {The
  Journal of chemical physics}\ }\textbf {\bibinfo {volume} {124}},\ \bibinfo
  {pages} {234104} (\bibinfo {year} {2006})}\BibitemShut {NoStop}%
\bibitem [{\citenamefont {Carr{\'e}}\ \emph {et~al.}(2007)\citenamefont
  {Carr{\'e}}, \citenamefont {Berthier}, \citenamefont {Horbach}, \citenamefont
  {Ispas},\ and\ \citenamefont {Kob}}]{carre2007amorphous}%
  \BibitemOpen
  \bibfield  {author} {\bibinfo {author} {\bibfnamefont {A.}~\bibnamefont
  {Carr{\'e}}}, \bibinfo {author} {\bibfnamefont {L.}~\bibnamefont {Berthier}},
  \bibinfo {author} {\bibfnamefont {J.}~\bibnamefont {Horbach}}, \bibinfo
  {author} {\bibfnamefont {S.}~\bibnamefont {Ispas}}, \ and\ \bibinfo {author}
  {\bibfnamefont {W.}~\bibnamefont {Kob}},\ }\href@noop {} {\bibfield
  {journal} {\bibinfo  {journal} {The Journal of chemical physics}\ }\textbf
  {\bibinfo {volume} {127}},\ \bibinfo {pages} {114512} (\bibinfo {year}
  {2007})}\BibitemShut {NoStop}%
\bibitem [{\citenamefont {Zahn}\ \emph {et~al.}(2002)\citenamefont {Zahn},
  \citenamefont {Schilling},\ and\ \citenamefont {Kast}}]{zahn2002enhancement}%
  \BibitemOpen
  \bibfield  {author} {\bibinfo {author} {\bibfnamefont {D.}~\bibnamefont
  {Zahn}}, \bibinfo {author} {\bibfnamefont {B.}~\bibnamefont {Schilling}}, \
  and\ \bibinfo {author} {\bibfnamefont {S.~M.}\ \bibnamefont {Kast}},\
  }\href@noop {} {\bibfield  {journal} {\bibinfo  {journal} {The Journal of
  Physical Chemistry B}\ }\textbf {\bibinfo {volume} {106}},\ \bibinfo {pages}
  {10725} (\bibinfo {year} {2002})}\BibitemShut {NoStop}%
\bibitem [{\citenamefont {Berendsen}\ \emph {et~al.}(1987)\citenamefont
  {Berendsen}, \citenamefont {Grigera},\ and\ \citenamefont
  {Straatsma}}]{doi:10.1021/j100308a038}%
  \BibitemOpen
  \bibfield  {author} {\bibinfo {author} {\bibfnamefont {H.~J.~C.}\
  \bibnamefont {Berendsen}}, \bibinfo {author} {\bibfnamefont {J.~R.}\
  \bibnamefont {Grigera}}, \ and\ \bibinfo {author} {\bibfnamefont {T.~P.}\
  \bibnamefont {Straatsma}},\ }\href {\doibase 10.1021/j100308a038} {\bibfield
  {journal} {\bibinfo  {journal} {The Journal of Physical Chemistry}\ }\textbf
  {\bibinfo {volume} {91}},\ \bibinfo {pages} {6269} (\bibinfo {year}
  {1987})},\ \Eprint
  {http://arxiv.org/abs/http://pubs.acs.org/doi/pdf/10.1021/j100308a038}
  {http://pubs.acs.org/doi/pdf/10.1021/j100308a038} \BibitemShut {NoStop}%
\bibitem [{\citenamefont {Kusaka}\ \emph {et~al.}(1998)\citenamefont {Kusaka},
  \citenamefont {Wang},\ and\ \citenamefont {Seinfeld}}]{kusaka_binary_1998}%
  \BibitemOpen
  \bibfield  {author} {\bibinfo {author} {\bibfnamefont {I.}~\bibnamefont
  {Kusaka}}, \bibinfo {author} {\bibfnamefont {Z.-G.}\ \bibnamefont {Wang}}, \
  and\ \bibinfo {author} {\bibfnamefont {J.~H.}\ \bibnamefont {Seinfeld}},\
  }\href {\doibase 10.1063/1.476097} {\bibfield  {journal} {\bibinfo  {journal}
  {The Journal of Chemical Physics}\ }\textbf {\bibinfo {volume} {108}},\
  \bibinfo {pages} {6829} (\bibinfo {year} {1998})}\BibitemShut {NoStop}%
\bibitem [{\citenamefont {Allen}\ and\ \citenamefont
  {Tildesley}(1987)}]{allen1987computer}%
  \BibitemOpen
  \bibfield  {author} {\bibinfo {author} {\bibfnamefont {M.~P.}\ \bibnamefont
  {Allen}}\ and\ \bibinfo {author} {\bibfnamefont {D.~J.}\ \bibnamefont
  {Tildesley}},\ }\href@noop {} {\emph {\bibinfo {title} {Computer simulation
  of liquids}}}\ (\bibinfo  {publisher} {Clarendon press Oxford},\ \bibinfo
  {year} {1987})\BibitemShut {NoStop}%
\bibitem [{Note2()}]{Note2}%
  \BibitemOpen
  \bibinfo {note} {In the case of the lamellar phases at low hydration levels,
  the studied time scales were so extended to observe a peculiar form of the
  `flying ice cube' issue~\cite {Harvey1997}. In this case, the energy of high
  frequency modes was transferred to the centers of mass of two sub-systems,
  formed by the absorbed fluid and the surfactant domains, respectively. Next,
  due to the conservation of {\protect \em total} momentum, the two subsystems
  moved ballistically in opposite directions, parallel to the planes of the
  lamellae. We cured this issue by periodically re-sampling the velocities of
  all atoms from the adequate Maxwell-Boltzmann distribution.}\BibitemShut
  {Stop}%
\bibitem [{\citenamefont {Plimpton}(1995)}]{citeulike:2835214}%
  \BibitemOpen
  \bibfield  {author} {\bibinfo {author} {\bibfnamefont {S.}~\bibnamefont
  {Plimpton}},\ }\href {\doibase 10.1006/jcph.1995.1039} {\bibfield  {journal}
  {\bibinfo  {journal} {Journal of Computational Physics}\ }\textbf {\bibinfo
  {volume} {117}},\ \bibinfo {pages} {1} (\bibinfo {year} {1995})}\BibitemShut
  {NoStop}%
\bibitem [{\citenamefont {Sunda}\ and\ \citenamefont
  {Venkatnathan}(2013)}]{sunda2013molecular}%
  \BibitemOpen
  \bibfield  {author} {\bibinfo {author} {\bibfnamefont {A.~P.}\ \bibnamefont
  {Sunda}}\ and\ \bibinfo {author} {\bibfnamefont {A.}~\bibnamefont
  {Venkatnathan}},\ }\href@noop {} {\bibfield  {journal} {\bibinfo  {journal}
  {Journal of Materials Chemistry A}\ }\textbf {\bibinfo {volume} {1}},\
  \bibinfo {pages} {557} (\bibinfo {year} {2013})}\BibitemShut {NoStop}%
\bibitem [{\citenamefont {Harvey}\ \emph {et~al.}(1997)\citenamefont {Harvey},
  \citenamefont {Tan},\ and\ \citenamefont {Iii}}]{Harvey1997}%
  \BibitemOpen
  \bibfield  {author} {\bibinfo {author} {\bibfnamefont {S.~C.}\ \bibnamefont
  {Harvey}}, \bibinfo {author} {\bibfnamefont {R.~K.}\ \bibnamefont {Tan}}, \
  and\ \bibinfo {author} {\bibfnamefont {T.~E.~C.}\ \bibnamefont {Iii}},\
  }\href@noop {} {\ \textbf {\bibinfo {volume} {19}},\ \bibinfo {pages} {726}
  (\bibinfo {year} {1997})}\BibitemShut {NoStop}%
\end{thebibliography}%
\end{document}